\documentclass[12pt]{article}
\usepackage[T1]{fontenc}
\usepackage{a4}
\usepackage{latexsym} 
\usepackage{amssymb}  
\usepackage{graphicx}
\usepackage{epsfig}
\begin{document}

\newcounter{zyxabstract}     
\newcounter{zyxrefers}        

\newcommand{\newabstract}
{\newpage\stepcounter{zyxabstract}\setcounter{equation}{0}
\setcounter{footnote}{0}}

\newcommand{\rlabel}[1]{\label{zyx\arabic{zyxabstract}#1}}
\newcommand{\rref}[1]{\ref{zyx\arabic{zyxabstract}#1}}

\renewenvironment{thebibliography}[1] 
{\section*{References}\setcounter{zyxrefers}{0}
\begin{list}{ [\arabic{zyxrefers}]}{\usecounter{zyxrefers}}}
{\end{list}}
\newenvironment{thebibliographynotitle}[1] 
{\setcounter{zyxrefers}{0}
\begin{list}{ [\arabic{zyxrefers}]}
{\usecounter{zyxrefers}\setlength{\itemsep}{-2mm}}}
{\end{list}}

\renewcommand{\bibitem}[1]{\item\rlabel{y#1}}
\renewcommand{\cite}[1]{[\rref{y#1}]}      
\newcommand{\citetwo}[2]{[\rref{y#1},\rref{y#2}]}
\newcommand{\citethree}[3]{[\rref{y#1},\rref{y#2},\rref{y#3}]}
\begin{titlepage}
\begin{flushright}
\small{
FZJ-IKP(TH)-2002-04\\
LU-TP 02-03\\
hep-ph/0201266
}
\end{flushright}

\vspace{1cm}

\begin{center}
{\huge\bf EFFECTIVE FIELD THEORIES\\[1cm] OF QCD}\\[1cm]
264. WE-Heraeus-Seminar\\
Physikzentrum Bad Honnef, Bad Honnef, Germany\\
November 26 --- 30, 2001\\[2cm]
{\bf Johan Bijnens}\\[0.3cm]
Department of Theoretical Physics 2, Lund University\\
S\"olvegatan 14A, S22362 Lund, Sweden\\[1cm]
{\bf Ulf-G. Mei\ss ner and Andreas Wirzba}\\[0.3cm]
{Forschungszentrum J\"ulich, Institut f\"ur Kernphysik (Theorie)\\ 
D-52425 J\"ulich, Germany}\\[2cm]
{\large ABSTRACT}
\end{center}
These are the proceedings of the workshop on ``Effective Field Theories
of QCD''
held at the Physikzentrum Bad Honnef of the
Deutsche Physikalische Gesellschaft, Bad Honnef, Germany from November 26 to
30, 2001. The workshop concentrated on Chiral Perturbation
Theory in its various settings, its relations with lattice QCD
and dispersion theory and the use of effective field theories
in systems with one, two or more nucleons as well as at high baryon densities.

Included are a short contribution per talk
and a listing of some review papers on the subject.

\end{titlepage}

\section{Introduction}
The use of effective field theory techniques is an ever growing approach
in various fields of  theoretical physics. 
Along with the continuing application of 
Chiral Perturbation Theory, Nuclear Effective Field Theory and, more
recently, QCD at high baryon density have been the focus of many
investigations. We therefore decided to organize the next topical workshop.
This meeting followed the series of workshops in Ringberg (Germany), 1988,
Dobog\'ok\"o (Hungary), 1991, Karreb\ae ksminde (Denmark), 1993,
Trento (Italy), 1996 and Bad Honnef (Germany), 1998.
All these workshops shared the same features,
about 50 participants, a fairly large amount of time devoted to discussions
rather than presentations and an intimate environment with lots of
discussion opportunities.

This meeting took place in late fall 2001 in the Physikzentrum Bad Honnef
in Bad Honnef, Germany and the funding provided by the 
WE--Heraeus--Stiftung allowed us to provide for the local
expenses for all participants and to support the travel of a fair amount of
participants. The WE-Heraeus foundation also provided
the administrative support for the workshop in the person of the
able secretary Heike Uebel. We extend our sincere gratitude to the WE-Heraeus
Stiftung for this support. We would also like to thank the staff of the
Physikzentrum for the excellent service given to us during the workshop
and last but not least the participants for making this an exciting
and lively meeting.

The meeting had 57 participants whose names, institutes and email addresses
are listed below. 48 of them presented results in presentations of various
lengths.
A short description of their contents and a list of the most relevant
references can be found below. As in the previous three of these workshops we
felt that this was more appropriate a framework than full-fledged proceedings.
Most results are or will soon be published and available on the archives,
so this way we can achieve speedy publication and avoid duplication of
results in the archives.

Below follows first the program, then the list of participants and
a subjective list of review papers, lectures and other proceedings
relevant to the subject of this workshop.
\\[1cm]
\hfill Johan Bijnens, Ulf-G. Mei\ss ner and Andreas Wirzba

\newpage
\section{The Program}
\begin{tabbing}
xx:xx \= A very very very long name \= \kill\\
{\bf Monday, November 26th, 2001}\\
{\it  Early Afternoon  Session}\\
 Chairperson: Johan   Bijnens    \>       \> 
{\bf  Chiral Perturbation Theory} \\

  14:00 \>    Ulf-G. Mei{\ss}ner /    \>  Introductory Remarks \\
        \>    Ernst Dreisigacker  \> \\
  14:20 \>    Barry R. Holstein   \>  Effective Field Theory and Gravity\\
         \>     (Amherst)\\
  14:55  \>   Jürg Gasser (Bern)  \> The Quark Condensate from $K_{e4}$
                                    Decays\\
  15:35  \>   Bastian Kubis  (Jülich) \> 
      Isospin Violation in Pion-Kaon
              Scattering\\
  16:00   \>                                Coffee\\
{\it   Late Afternoon Session}\\
  Chairperson: Andreas Wirzba  \>\>  {\bf Chiral Perturbation Theory}\\
       16:30    \>      Joaquim Prades  (Granada)\>
    Matching the Elektroweak Penguins      $Q_7$ and $Q_8$\\\>\> at NLO\\
       17:05 \>         Paul Büttiker  (Jülich) \>  
    Chiral and Dispersive Description  of Pion Kaon\\\>\> Scattering\\
       17:30 \>         Elisabetta Pallante  (Trieste)\>
  Kaon Decays on and off the Lattice\\
       17:55 \>         Pere Talavera   (Marseille)\>
 On the Light Meson Formfactors\\
       18:20  \>                          End of Session\\
       18:30  \>                              Dinner\\[0.5cm]

{\bf  Tuesday, November 27th, 2001}\\
{\it  Early Morning Session}\\
 Chairperson: Ulf-G. Mei{\ss}ner\>\>{\bf Chiral Perturbation Theory}\\
09:00  \>    Gerhard Ecker (Wien) \>  
Hadronic Vacuum Polarization\\
09:40  \>    Akaki Rusetsky (Bern) \> 
Hadronic Atoms in Effective Field Theories\\
10:20   \>   Santiago Peris (Barcelona)\>
Large-$N_c$ QCD and Weak Matrix      Elements\\
10:45   \>                             Coffee\\
{\it   Late Morning Session}\\
 Chairperson: Ulf-G     Mei{\ss}ner \>\>{\bf Chiral Perturbation Theory}\\
11:15 \>     Eugene Golowich  (Amherst)  \>   
Determination of $\langle (\pi\pi)_{I=2}|Q_{7,8}|K^0\rangle$ 
in the\\\>\> Chiral Limit\\
11:50 \>     Luca Girlanda (Padova)\>
 Analysis and Interpretation of New Low-Energy\\\>\>
 Pion Pion Scattering Data\\
12:15  \>    Jan Stern (Orsay) \> 
    New Approach to Three-Flavour Chiral Dynamics\\
12:55 \>                           End of Session\\
13:00 \>                                Lunch\\
{\it Early Afternoon Session}\\
Chairperson: Johan  Bijnens\>\>{\bf Chiral Perturbation Theory}\\
      14:20 \>     Marc Knecht    (Marseille) \>     
  Low-Energy Physics of the Standard  Model and\\\>\> Large-$N_c$ QCD\\
      14:50 \>     Andreas Nyffeler \>   
 Hadronic Light-by-Light Scattering  Corrections to the\\
  \>  (Marseille)\> Muon g-2: the
                                        Pion-Pole Contribution\\
      15:10  \>    Sebastian Descotes \>   
 $N_f$-Sensitivity of Chiral Dynamics\\
 \> (Southampton)\>\\
      15:35 \>     Hartmut Wittig  (Liverpool)   \>  
 Effective Chiral Lagrangian from   Simulations of
\\\>\> Partially Quenched QCD\\
      16:00 \>                               Coffee\\
{\it  Late Afternoon Session}\\
Chairperson: Johan Bijnens \>\>{\bf Chiral Perturbation Theory}\\
      16:30 \>     Vincenzo Cirigliano  (Wien) \>   
Radiative Corrections to $K_{l3}$ Decays\\
      16:55 \>     Roland Kaiser (San  Diego) \> 
 Light Quark Mass Ratios from Large  $N_c$ Chiral
\\\>\> Perturbation Theory\\
      17:20\>      Frederik Persson   (Lund)    \>  
 Recalculation and Reanalysis of $K\to 3\pi$\\
      17:55 \>     Bugra Borasoy  (München)    \>     
Decays of the $\eta'$ in Chiral  Perturbation Theory\\
      18:20 \>                           End of Session\\
      18:30 \>                               Dinner\\[0.5cm]
{\bf  Wednesday, November 28th, 2001}\\
{\it Early Morning Session}\\
  Chairperson: Andreas Wirzba \>\>{\bf       Chiral Nucleon Dynamics}\\
      09:00 \>     Dieter Drechsel  \>    The Polarizability of the Nucleon:\\
            \>     (Mainz)       \>       The View of Dispersion Relations\\
      09:35 \>     Pavel Pobylitsa  \> 
 Chiral Dynamics and Large-Nc Baryons:   Model-\\
             \>      (Bochum) \> Independent Results for Parton
                                      Distributions\\

      10:05\>      Thomas R. Hemmert\> 
Chiral EFT with Explicit Spin 3/2 Fields:\\
           \>      (München)        \>   A Status Report\\
      10:40 \>                               Coffee\\
{\it   Late Morning Session}\\
 Chairperson:     Andreas Wirzba \>\>{\bf      Chiral Nucleon Dynamics}\\
      11:15 \>     Matthias Frink  (Jülich)  \>  
Analysis of the Pion-Kaon Sigma Term\\
      11:40 \>     Maxim V. Polyakov \>   New Soft Pion Theorem for Hard Near\\
             \>     (Bochum)       \>       Threshold Pion Production\\
      12:05 \>     Eulogio Oset    \>  
   Chiral Dynamics in Meson-Baryon, N-N\\
             \>     (Valencia)    \>        and Weak LambdaN-NN Interactions\\
      12:45 \>                           End of Session\\
      13:00 \>                                Lunch\\
{\it Early Afternoon Session}\\
 Chairperson: Ulf-G.      Mei{\ss}ner\>\>{\bf Chiral Few-Nucleon Dynamics}\\
      14:30 \>     Silas R. Beane  (Seattle) \>     Perturbative Theories of Nuclear  Forces\\
      15:10 \>     Joan Soto    (Barcelona)   \>
   Renormalizing the Lippmann-Schwinger\\
            \>       \>      Equation for the OPE Potential\\
      15:35 \>   Harald W.     Grie{\ss}hammer   \>  
    Very Low Energy Deuteron Compton  Scattering and\\
             \>      (München)     \>      
  Dispersive Effects in  Nucleon Polarisabilities\\
      16:00 \>                               Coffee\\
{\it  Late Afternoon Session}\\
Chairperson: Ulf-G.      Mei{\ss}ner\>\>{\bf Chiral Few-Nucleon Dynamics}\\
      16:30\>      Hermann Krebs  (Jülich)  \>   
  Pion Electroproduction on the\\
            \>            \>      Deuteron Near the Threshold\\
      16:55 \>     Bira van Kolck   (Tucson)    \> 
 One Two Three ... Infinity: Nucleons in EFT\\
      17:35 \>     Hans-Werner Hammer \>  Few-Body Physics in Effective Field\\
            \>     (Ohio)          \>     Theory\\
      18:15 \>                           End of Session\\
      18:30 \>                               Dinner\\[0.5cm]
{\bf Thursday, November 29th, 2001}\\
{\it  Early Morning Session}\\
 Chairperson: Johan     Bijnens\>\>{\bf
             Chiral Few- and Many-Nucleon Dynamics}\\
      09:00 \>     Walter Glöckle \>   Applications of Chiral Nuclear Forces\\
              \>    (Bochum)     \>        to Few-Nucleon Systems\\

      09:40 \>  Evgeny Epelbaum  \>  Nuclear Forces from Chiral EFT: Recent\\
              \>    (Jülich)   \>          Developments\\
      10:20 \>     Norbert Kaiser  (München) \>   
 Chiral Dynamics of Nuclear Matter\\
      10:55 \>                               Coffee\\
{\it   Late Morning Session}\\
  Chairperson: Johan Bijnens\>\>{\bf Chiral Many-Body Dynamics and High
     Density}\\
      11:25 \>     José A. Oller  (Jülich)   \>
    In-Medium ChPT beyond the Mean-Field\\
             \>      \>        Approach \\
      12:05 \>     Mark Alford   (Glasgow)  \>
     Color Superconductivity in  High-Density QCD\\
      12:45 \>                           End of Session\\
      13:00 \>                                Lunch\\
{\it Early Afternoon Session}\\
Chairperson:     Andreas Wirzba \>\>{\bf     High-Density QCD}\\
      14:30 \>     Krishna Rajagopal\>   Crystalline Color Superconductivity\\
             \>    (Cambridge)\> \\
      15:10 \>     Dirk Rischke    (Frankfurt)  \>    The Prefactor of the
  Color-Superconducting Gap\\
      15:50 \>                               Coffee\\
{\it Late Afternoon  Session}\\
Chairperson: Andreas Wirzba\>\>
{\bf Effective Theories of High-Density QCD}\\
     \>\> {\bf and Lattice QCD}\\
      16:20 \>     Roberto Casalbuoni \>
 Effective Theories for QCD at High  Density\\
           \>       (Firenze)   \>       \\
      17:00 \>     Ismail Zahed (Stony Brook) \>Dense QCD\\
      17:40 \>     Harald Markum  (Wien)    \>
    Color Superfluidity in Two Color QCD\\
              \>   \>      on the Lattice\\
      18:05 \>                           End of Session\\
      18:30 \>                               Dinner\\[0.5cm]
{\bf Friday, November 30th, 2001}\\
{\it Early Morning Session}\\
Chairperson: Ulf-G.    Mei{\ss}ner\>\>{\bf  Lattice QCD}\\
      09:00 \>    Martin Savage (Seattle) \>  Baryons in Quenched Chiral
                Perturbation Theory\\
      09:40 \>    Tilo Wettig (New Haven) \> Chiral Symmetry and the Spectrum
   of the QCD\\\>\> Dirac Operator\\
      10:15 \>    Christof Gattringer \>    
  Calorons Near the QCD Phase Transition\\
             \>   (Regensburg)       \>   \\
      10:40 \>                               Coffee\\
{\it  Late Morning Session}\\
 Chairperson: Ulf-G.     Mei{\ss}ner\>\>{\bf Lattice QCD}\\
      11:10 \>    Klaus Schilling \>   
$\eta^\prime$ Mass from QCD on the Lattice\\
              \>  (Wuppertal)\> \\
      11:45 \>    C.-J. David Lin  \>       
  Non-leptonic Kaon Decays from  Lattice QCD\\
             \>   (Southampton)    \>  \\
      12:10 \>    Johan Bijnens   \>          Farewell\\
     12:20 \>                      Lunch and End of Workshop\\
\end{tabbing}

\section{Participants and their email}

\begin{tabbing}
A very long name\=a very long institute\=email\kill
M. Alford  \>Glasgow \>alford@physics.gla.ac.uk\\
S. Beane  \>Seattle \>sbeane@phys.washington.edu\\
V. Bernard \>Strasbourg\> bernard@lpt6.u-strasbg.fr\\
J. Bijnens \>Lund\>   bijnens@thep.lu.se    \\
B. Borasoy \>TU M\"unchen\>borasoy@physik.tu-muenchen.de     \\
P. Buettiker \>FZ J\"ulich\>  p.buettiker@fz-juelich.de    \\
R. Casalbuoni \>Florence\>  casalbuoni@fi.infn.it    \\
V. Cirigliano \>Wien\> vincenzo@doppler.thp.univie.ac.at  \\
S. Descotes \>Southampton\>  sdg@hep.phys.soton.ac.uk \\
D. Drechsel \>Mainz\>   drechsel@kph.uni-mainz.de  \\
G. Ecker \>Wien\> ecker@thp.univie.ac.at         \\
E. Epelbaum \>FZ J\"ulich\> evgenie.epelbaum@tp2.ruhr-uni-bochum.de   \\
M. Frink \>FZ J\"ulich\> m.frink@fz-juelich.de  \\
J. Gasser \>Bern\>  gasser@itp.unibe.ch       \\
C. Gattringer \>Regensburg\>christof.gattringer@physik.uni-regensburg.de \\
L. Girlanda \>Padova\> girlanda@pd.infn.it     \\
W. Gl\"ockle \> Bochum \> walter.gloeckle@tp2.ruhr-uni-bochum.de\\
E. Golowich \>Amherst\> golowich@physics.umass.edu         \\
H. Grie{\ss}hammer \>M\"unchen\>hgrie@physik.tu-muenchen.de  \\
H. Hammer \>Ohio State\>hammer@mps.ohio-state.edu   \\
T. Hemmert \>TU M\"unchen\>  themmert@physik.tu-muenchen.de     \\
I. Hip \>Wuppertal\> hip@theorie.physik.uni-wuppertal.de  \\
B. Holstein \>Amherst\> holstein@physics.umass.edu         \\
N. Kaiser \>TU M\"unchen\> nkaiser@physik.tu-muenchen.de         \\
R. Kaiser \>San Diego\> kaiser@pauli.ucsd.edu        \\
M. Knecht \>Marseille\>knecht@cpt.univ-mrs.fr    \\
H. Krebs \>FZ J\"ulich\>h.krebs@fz-juelich.de     \\
B. Kubis \>FZ J\"ulich\>b.kubis@fz-juelich.de     \\
C.-J. David Lin \> Southampton \> dlin@hep.phys.soton.ac.uk \\ 
H. Markum \>TU Wien\> markum@tuwien.ac.at   \\
U.-G. Mei{\ss}ner  \>FZ J\"ulich\> u.meissner@fz-juelich.de   \\
H. Neufeld \>Wien\>  neufeld@merlin.pap.univie.ac.at      \\
A. Nyffeler \>Marseille\> nyffeler@cpt.univ-mrs.fr      \\
J. Oller \>FZ J\"ulich\> oller@um.es     \\
E. Oset \>Valencia\> oset@condor1.ific.uv.es     \\
E. Pallante \>S.I.S.S.A.\> pallante@he.sissa.it  \\
S. Peris \>Barcelona\> peris@ifae.es    \\
F. Persson \>Lund\>   fredrik.persson@thep.lu.se    \\
H. Petry \>Bonn\> petry@itkp.uni-bonn.de        \\
L. Platter \>FZ J\"ulich\> l.platter@fz-juelich.de  \\
P. Pobylitsa \>Bochum\> pavel.pobylitsa@tp2.ruhr-uni-bochum.de      \\
M. Polyakov \>Bochum\> maximp@tp2.ruhr-uni-bochum.de      \\
J. Prades \>Granada\> prades@ugr.es     \\
K. Rajagopal \>MIT\> krishna@ctp.mit.edu   \\
D. Rischke \>Frankfurt\> drischke@th.physik.uni-frankfurt.de     \\
A. Rusetsky \>Bern\> rusetsky@itp.unibe.ch     \\
M. Sainio \>Helsinki\>sainio@phcu.helsinki.fi     \\
M. Savage \>Seattle\> savage@phys.washington.edu     \\
K. Schilling \>Wuppertal\>schillin@wpts0.physik.uni-wuppertal.de    \\
J. Soto \>Barcelona\> soto@ecm.ub.es     \\
J. Stern \>Orsay\> stern@ipno.in2p3.fr        \\
P. Talavera \>Barcelona\> pere@ecm.ub.es \\
U. van Kolck \>Tucson\>vankolck@physics.arizona.edu \\
T. Wettig \>Yale\>  tilo.wettig@yale.edu     \\
A. Wirzba \> FZ J\"ulich\>a.wirzba@fz-juelich.de\\
H. Wittig \>Liverpool\>  wittig@amtp.liv.ac.uk    \\
I. Zahed \>Stony Brook\>zahed@zahed.physics.sunysb.edu \\
\end{tabbing}
\newabstract
\section{A short guide to review literature}

Chiral Perturbation Theory grew out of current algebra, and it  soon
was realized that certain terms beyond the lowest order were also uniquely
defined. This early work and references to earlier review papers can be found
in \cite{1}. Weinberg then proposed a systematic method in \cite{2},
later systematized
and extended to use the external field method in the classic papers
by Gasser and Leutwyler \cite{3},\cite{4}, which, according to Howard Georgi,
everybody should put under his/her pillow before he/she goes to sleep.
The field has since then extended a lot
and relatively recent review papers are: Ref.\cite{5} with an emphasis on the
anomalous sector, Ref.\cite{6} giving a general overview over the vast field
of applications in  various areas of physics, Ref.\cite{7} on mesons and
baryons,
and Ref.\cite{8} on baryons and multibaryon processes. This is updated
in Ref.\cite{8a}.
In
addition there are books
by Georgi\cite{9}, which, however, does not cover the standard approach,
including the terms in the lagrangian at higher order and
a more recent one by Donoghue, Golowich and Holstein\cite{10}.

There are also the lectures available on the archives by H.~Leutwyler
\cite{10b}
E.~de~Rafael \cite{11}, A.~Pich \cite{12b}, G. Ecker \cite{12c}
as well as numerous others (the single nucleon sector is covered in
most detail in \cite{12d}).
The references to the previous
meetings are \cite{13},\cite{14},\cite{15},\cite{15b},\cite{15c}.
There are also the proceedings of the Chiral Dynamics
meetings at MIT (1994) \cite{16}, in Mainz (1997) \cite{16b}
and at Jefferson Lab \cite{16c}.
The DA$\Phi$NE
handbook \cite{17} also contains useful overviews. The few-nucleon sector
is covered in the proceedings of the Caltech and Seattle workshops \cite{18},
\cite{19}, respectively,
in the reviews by van Kolck \cite{20} and by Beane, Bedaque, Haxton,
Philips and Savage
\cite{21}. There also exist reviews on the novel developments in dense QCD,
see Refs.\cite{Reviews}. A recent condensed overview can be found in
\cite{LP01}.

\newabstract 
\begin{center}
{\bf Gravity and Effective Field Theory}\\
Barry R. Holstein, Department of Physics, University of Massachusetts\\
and IKP-Theorie, Forschungszentrum J\"{u}lich
\end{center}

We use effective field theory techniques to examine the quantum corrections 
   to the gravitational 
   metrics of charged particles, with and without spin.  Gravity couples 
to the energy momentum tensor $T_{\mu\nu}$ of a particle\cite{wb}. We 
   calculate the energy momentum tensor in a power series in $\alpha$, using 
   usual Feynman diagram techniques.   The results are expressed in terms of form factors $F(q^2)$
   of the various allowed Lorentz structures in the matrix element of $T_{\mu\nu}$.
 Normally, form factors can be expanded in a power series in $q^2$
   around
   $q^2=0$, with the coefficients being related to the structure of the particle. 
However, in momentum space 
   the masslessness of the photon implies
   the presence of nonanalytic pieces $\sim\sqrt{-q^2},q^2\log -q^2$,
   etc. in the form factors of the energy-momentum tensor.  
By transforming to coordinate space we show how the
   former reproduces the classical non-linear terms of the 
Reissner-Nordstr\"{o}m\cite{rn}
   and Kerr-Newman\cite{kn}
   metrics or order $G\alpha/r^2$ while the latter can be interpreted as quantum 
corrections to these
   metrics, of order $G\alpha\hbar/mr^3$.

\newabstract 

\begin{center}
{\large{\bf The quark condensate from $K_{e4}$ decays}}\\[.5cm]
G.~Colangelo$^1$,
{\bf {J.~Gasser}$^2$} and  
H.~Leutwyler$^2$\\[.3cm]
\footnotesize{\begin{tabular}{c}
$^1\,$Institute for Theoretical Physics, University of 
Z\"urich\\
Winterthurerstr. 190, CH-8057 Z\"urich, Switzerland\\[.3cm]
$^2\,$Institute for Theoretical Physics, University of 
Bern\\
Sidlerstr. 5, CH-3012 Bern, Switzerland
\end{tabular}  }

\end{center}

Roy equations \cite{roy} combined with chiral symmetry allow one to predict
\cite{cgl}  the
low-energy behavior of the $\pi\pi$ scattering amplitude with high
precision. The prediction is based on the hypothesis that the quark
condensate is the leading order parameter of spontaneously broken
chiral symmetry. This has been questioned  in Refs. \citetwo{p1}{p2},
where it is
emphasized that experimental evidence for this scenario is not
available. The recent result from the high statistics $K_{e4}$
experiment E865 at Brookhaven \cite{e865} closes this gap, because those data
allow one to determine the size of the leading term in the quark mass
expansion of the pion mass. E865 confirms \cite{e865} the hypothesis that
underlies our prediction of the  $\pi\pi$ $S$-wave scattering lengths:
 more than 94\% of the pion mass stems from the quark condensate. The
generalized framework of $SU(2)\times SU(2)$ chiral perturbation theory
developed in \cite{p1}, that allows for a small or vanishing condensate,
can therefore be dismissed.\\[.5cm]
{\em Note added}: In Ref. \cite{dgfs}, the E865 data \cite{e865} have been 
analyzed without the use of chiral symmetry constraints.

\newabstract 

\begin{center}
{\large\bf Isospin violation in low--energy pion--kaon scattering}\\[0.5cm]
{Bastian Kubis}\\[0.3cm]
Forschungszentrum J\"ulich, Institut f\"ur Kernphysik (Theorie), \\
D--52425 J\"ulich, Germany\\[0.3cm]
\end{center}

In order to extract the two pion--kaon S--wave scattering lengths from 
$\pi K$ bound state experiments~\cite{DIRAC},
the measurable observables have to be related to 
combinations of these scattering lengths  by modified Deser formulae
that include next--to--leading order effects in isospin breaking.
In particular, these involve the isospin violating corrections in the regular parts of
the scattering amplitudes $\pi^- K^+ \to \pi^0 K^0 \,\, (\pi^- K^+)$ at threshold.
We have evaluated these corrections up to one--loop order, including
leading effects both in the light quark mass difference $m_u-m_d$ and the electromagnetic
fine structure constant $\alpha = e^2/4\pi$ \cite{KM} (see also~\cite{Nehme}).
The numerical results are
\begin{eqnarray}
&& \hspace{-0.8cm}
a_0\left(\pi^-K^+ \to \pi^0K^0\right) \,= 
\\ &&  \hspace{-0.25cm}
- \sqrt{2} \, a_0^- \biggl\{ \,
\underbrace{( 1 \pm 0.8\% )}_{{\cal O}(p^4)} 
\,+\, \underbrace{\!\phantom{(} 0.8\% \phantom{)}\!}_{{\cal O}(e^2)} 
\,+\!\! \underbrace{\!\phantom{(} 0.5\% \phantom{)}\!}_{{\cal O}(m_u-m_d)} 
\!\!-\, \underbrace{( 0.8 \pm 0.7 )\%}_{{\cal O}(p^2e^2)} 
\,+\, \underbrace{( 0.7 \pm 0.2 )\%}_{{\cal O}(p^2(m_u-m_d))} \,
\biggr\} ~, \rlabel{00corr} \nonumber \\
&& \hspace{-0.8cm}
a_0\left(\pi^-K^+\to\pi^-K^+\right) \,= 
\\ && \qquad \qquad \quad
\left(a_0^- + a_0^+\right) \biggl\{ \,
\underbrace{( 1 \pm 16.1\% )}_{{\cal O}(p^4)} 
\,+   \underbrace{\!\phantom{(} 1.2\% \phantom{)}\!}_{{\cal O}(e^2)} 
  -\, \underbrace{( 0.3 \pm 3.2 )\%}_{{\cal O}(p^2e^2)} 
\,+\!\!\!\!\!\!
      \underbrace{\!\phantom{(} 0.2\% \phantom{)}\!}_{{\cal O}(p^2(m_u-m_d))} \!\!\!\!\!\!
\biggr\} ~. \nonumber \rlabel{-+corr}
\end{eqnarray}

The isospin symmetric one--loop representation allows for a very precise 
prediction of the isovector scattering length $a_0^-$ that  enters the 
pion--kaon atom lifetime formula. 
Isospin breaking corrections for the charged--to--neutral channel give rise 
to a small shift of about one percent, and only a slight increase of the error range.
The uncertainty in a chiral prediction for the $2S-2P$ energy level shift
is however dominated by the large uncertainty in the
isoscalar scattering length $a_0^+$. 
Isospin breaking corrections in the relevant charged--to--charged channel are again small,
though with a larger error range, which still allow for a sufficiently accurate
extraction of the combination $a_0^-+a_0^+$ from experiment.

\newabstract 

\begin{center}
{\large\bf Matching the Electromagnetic Penguins
  $Q_7$ and $Q_8$}\\[0.5cm]
Johan Bijnens$^1$, Elvira G\'amiz$^2$, and 
{\bf Joaquim Prades}$^2$ \\[0.3cm]
$^1$Department Theor. Phys. 2, Lund University,\\
S\"olvegatan 14A, S 22362 Lund, Sweden\\[0.3cm]
$^2$
Centro Andaluz de F\'{\i}sica de las Part\'{\i}culas
Elementales (CAFPE) and \\
Departamento de F\'{\i}sica Te\'orica y del Cosmos, Universidad 
de Granada\\ Campus de Fuente Nueva, E-18002 Granada, Spain\\[0.3cm]
\end{center}

This talk describes the work done together with Hans Bijnens
and Elvira G\'amiz in \cite{BGP01}. 
Exact analytical expressions for the $\Delta S=1$ coupling Im $G_E$
in terms of observable spectral functions are given.
This coupling determines the size of  the $\Delta I=3/2$
contribution to $\varepsilon'$. We show analytically how the 
scheme and the scale dependences vanish to all orders in $1/N_c$ 
and NLO in $\alpha_S$ explictly both for $Q_7$ and $Q_8$. 
Numerical results are derived for both $Q_7$ and $Q_8$ from the
$\tau$-data and known results on the scalar spectral functions.
In particular, we study the effect of all higher dimension operators.
The coefficients of the leading operators in the OPE of the needed 
correlators are derived to NLO in $\alpha_S$. 
Related work can be found in the recent papers
\citethree{CDG}{NPR}{NAR}.

\newabstract 

\begin{center}
{\large\bf Chiral and Dispersive Description of $\mathbf{\pi K}$
           Scattering}\\[0.5cm]
B.~Ananthanarayan$^1$, {\bf P. B\"uttiker}$^2$ and B.~Moussallam$^3$\\[0.3cm]
$^1$CTS, Indian Institute of Science,
Bangalore~560~012, India\\[0.5mm]
$^2$IKP (Th), Forschungszentrum J\"ulich,
52425 J\"ulich, Germany \\[0.5mm]
$^3$IPN, Universit{\'e} Paris--Sud,
91406 Orsay C{\'e}dex, France\\[0.3cm]
\end{center}
ChPT provides the low energy effective theory of
the standard model that describes the interactions involving hadronic
degrees of freedom.
It requires the introduction of additional low energy
constants (LEC) at each order of the loop--expansion. In $SU(3)$ at
next--to--leading order, ten LECs $L^r_i(\mu)$ enter the calculations. 
Recently, there has been a
revival in interest in the constant $L_4^r$ as it
reflects the dependence of the decay constant $F_\pi$ in the chiral limit
on the number of chiral quarks \cite{fpidep}.
However, these quantities cannot
be fixed by chiral symmetry only; experimental information is
needed to pin them down. Using the $K_{l4}$ form
factors and $\pi\pi$ sum rules provide accurate estimates for some of the
LECs (e.g. $L^r_1, L^r_2, L^r_3$) but leave the large $N_c$--suppressed LEC
$L^r_4$  essentially undetermined. This is due to the fact that in such
processes, contributions from $L^r_4$ accidentally are suppressed.
Such a suppression does not occur in $\pi K$
scattering \cite{BKM}. Therefore this
process is suitable for a more reliable determination of $L^r_4$ and also
provides estimates for $L^r_1, L^r_2$ and $L^r_3$. To fix these
quantities we use dispersion relations \cite{ABM}, relying on analyticity, 
unitarity and crossing symmetry only, which are the suitable tools for a
comparison of experimental data with ChPT. By combining fixed--$t$
and hyperbolic dispersion relations for $T^+(s,t,u)$ and $T^-(s,t,u)$
and rewriting the chiral amplitudes, we find an appropriate matching of the
chiral and the dispersive representations of these amplitudes.
Saturating the dispersive integrals with the available data,
we find the following estimates (renormalization--scale $\mu=m_\rho$):
$$
   L_1^r = 0.84 \pm 0.15,\, L_2^r = 1.36 \pm 0.13,\,
   L_3^r = -3.65 \pm 0.45,\,L_4^r = 0.22 \pm 0.30
$$
yielding an estimate for $L^r_4$ with improved reliability.

\newabstract 

\begin{center}
{\large\bf Kaon decays on and off the lattice  }\\[0.5cm]
Elisabetta Pallante \\[0.3cm]
S.I.S.S.A., Via Beirut 2-4, 34014 Trieste, Italy\\[0.3cm]
\end{center}

The theoretical prediction of kaon-decay amplitudes, along with 
the determination of the $\Delta I=1/2$ ratio and the CP-violating parameter 
$\varepsilon'/\varepsilon$, still remains a challenging problem because of the 
crucial interplay between long- and short-distance contributions.
Relevant issues are the calculation of final state interactions (FSI), 
especially in the $I=0$ channel, and the consequences of implementing 
(partial) 
quenching in a lattice determination of $K\to\pi\pi$ or $K\to\pi$ amplitudes.

A Standard Model (SM) prediction of $\varepsilon'/\varepsilon$ has been 
derived in \cite{SM}, where FSI effects \cite{FSI} in the $I=0,2$ channels 
have been 
included by means of ChPT, $1/N_c$ expansion and the Omn\`es resummation.
The SM prediction ${\mbox{Re}}(\varepsilon'/\varepsilon ) = (17\pm 9)
\cdot 10^{-4}$ is in good agreement with the present experimental value
${\mbox{Re}}(\varepsilon'/\varepsilon ) = (17.2\pm 1.8)\cdot 10^{-4}$, 
although still affected by a large uncertainty which is dominated by the 
large-$N_c$ approximation and the uncertainty in the determination of
 the strange quark mass.

The calculation of weak matrix elements of strong penguin operators 
is implemented on the lattice within the (partially) quenched approximation,
with $K$ valence and ghost quarks and $N$ sea quarks ($N=0$ in the fully 
quenched limit).
It has been shown in \cite{PENGUIN} how partial quenching modifies 
the transformatin properties of QCD penguin operators under chiral symmetry.
In the (P)QChPT realization of $(8_L,1_R)$ penguin operators, new 
$(8_L,8_R)$ operators appear at leading order in the chiral expansion 
as pure quenching artifacts. It is crucial to uncover the size 
of the contamination due to the new operators \cite{LAT}, or alternatively
develop a strategy which removes those effects {\em ab initio}. This issue
is currently under study.

\newabstract 

\begin{center}
{\large\bf On the Light Meson Formfactors}\\[0.5cm]
J. Bijnens$^1$ and {\bf P. Talavera}$^2$\\[0.3cm]
$^1$Dep. Theor. Phys. 2, Lund University,\\
S\"olvegatan 14A, S22362 Lund, Sweden\\[0.3cm]
$^2$ 
    Departament d'Estructura i
    Constituents de la Materia\\
Universitat de Barcelona,
    Diagonal 647,
    Barcelona E-08028 Spain
\end{center}

We show the next-to-next-to-leading order (NNLO) results \cite{BP} on   
the pseudoscalar
vector formfactors in the three flavour case.      
Our results are confronted with earlier similar works in
the NLO case \cite{GL} and the NNLO in
the case of two flavours \citetwo{GM}{BCT}.                     

We discuss all the assumptions besides chiral symmetry that come
in the calculation, special attention is devoted to the large-N$_c$  
arguments. We update the value for the three
flavour ${\cal O}(p^4)$  parameters, $L_i^r,~i=1,\ldots,8$,
 \cite{ABT2} and
in addition we determine the low-energy constant $L_9^r$. 
Together with this we show also
the ratio $m_u/m_d$ for the quark-masses \cite{ABT4} as well
as the dependence on $m_s/\hat m$.

Finally we discuss some ad hoc changes in the values of $L_4^r$ and $L_6^r$
to have a ``well
behaved chiral series'' for the pseudoscalar masses.
This suggests that values for $L_4^r$ and
$L_6^r$ at the edge of the range given by
large-N$_c$ can satisfy the
requirement ${\cal O}(p^2) > {\cal O}(p^4) > {\cal O}(p^6)$ for all
existing calculated processes.

\newabstract 

\begin{center}
{\large\bf Hadronic Vacuum Polarization}\\[0.5cm]
Gerhard Ecker \\[0.3cm]
Institut f\"ur Theoretische Physik, Universit\"at Wien\\
Boltzmanngasse 5, A-1090 Vienna, Austria\\[1cm]
\end{center}

\noindent  
The hadronic vacuum polarization is an important ingredient for the
calculation of the anomalous magnetic moment of leptons and of the
running fine structure constant. For the muon anomalous magnetic
moment, the two-pion intermediate state provides the dominant
contribution. To incorporate the precise experimental results for
$\tau^- \to \pi^0 \pi^- \nu_\tau$ in the determination of hadronic
vacuum polarization, the leading isospin-violating and electromagnetic
corrections for this decay at low energies were calculated
\cite{CEN1}. The
corrections are small but relevant at the level of precision necessary
for a comparison between the standard model prediction for the muon
anomalous magnetic moment and experiment. More generally, the
corrections account for the main part of the systematic differences
between the measured form factors in $\tau^- \to \pi^0 \pi^- \nu_\tau$
and $e^+ e^- \to \pi^+ \pi^-$ at low to intermediate energies. 

In the calculation of the running fine structure constant entering the
analysis of electroweak precision measurements, a major theoretical
uncertainty is due to the four-pion intermediate state. The first
complete calculation of $e^+ e^- \to 4 \pi$ to $O(p^4)$ in chiral
perturbation theory is presented \cite{EU1}. Although the chiral 
amplitude cannot 
be used directly in the region of main interest ($E_{\rm cms} > 1$ 
GeV) it exhibits the correct low-energy structure to be continued 
into the resonance region. Important additional contributions 
occuring at $O(p^6)$ due to $\omega$ and $a_1$ exchange have been
included. Cross sections for the final states $2\pi^0 \pi^+ \pi^-$ 
and $2\pi^+ 2\pi^-$ are displayed for $E_{\rm cms} \le 1$ GeV.

\newabstract 

\begin{center}
{\large\bf Splitting Electromagnetic and Strong Interactions}\\[0.5cm]
J\"{u}rg Gasser, Irma Mgeladze, {\bf Akaki Rusetsky} and 
Ignazio Scimemi\\[0.3cm]
ITP, University of Bern, Sidlerstrasse 5, 3012 Bern, Switzerland\\[0.3cm]
\end{center}

In order to fully exploit the high-precision experimental data on
hadronic atoms which is available at present and which will be provided in
the future, it is
imperative to design a theoretical framework that describes this  type
of bound systems to an accuracy that
matches the experimental precision~\cite{Atoms}. 
The Deser-type relations which are then used  to extract the ``purely 
strong''
scattering lengths from the measured values of the strong energy shift 
in the ground state, and the partial decay width into 
hadronic channels, contain isospin-breaking corrections which depend on  
the fine structure constant $\alpha$ and on
the  quark mass difference $m_d-m_u$. The following question
 arises: The scattering lengths are calculated in QCD at $m_u=m_d$.
How is this theory related to a framework that includes
isospin violating effects,  $SU(3)_C \times U(1)_{em}$?
[Note that, in view of the announced accuracy $0.3\,\%$ 
in the future measurement of the energy shift of the $\pi^-p$ atom by  
 the Pionic Hydrogen collaboration at PSI, the question of the precise
definition of the pure QCD limit is not an academic one.]

The answer is the following~\cite{Splitting}:
In two-flavour QCD, the quark mass $m_u=m_d$ and strong coupling
constant may be  fixed from the requirement that the 
pion and nucleon masses are
equal - by convention - to the experimental values of the charged pion 
and the proton masses, respectively. In the effective theory 
of $SU(3)_C\times U(1)_{em}$, the values of the strong LECs stay 
put. 
With this definition, the isopin-breaking corrections to the bound-state
characteristics can be
calculated in a general theory of hadronic atoms~\cite{Atoms},
which has been successfully applied
to the calculation of the 
$\pi^+\pi^-$ atom decay width~\cite{Atoms}, and to the energy shift of the
$\pi^-p$ atom~\cite{Pimp}.

\newabstract 

\begin{center}
{\large\bf Large-$N_c$ QCD and weak matrix elements}\\[0.5cm]
S. Peris\\[0.3cm]
Grup de Fisica Teorica and IFAE, Univ. Autonoma de Barcelona,\\
08193 Bellaterra (Barcelona), Spain\\[0.3cm]
\end{center}

The use of large $N_c$ for the calculation of matrix elements of
electroweak operators was pioneered in the work of
\cite{Bardeen:1987uz}. With time, those ideas have evolved into a
new approach, the Minimal Hadronic Approximation  (MHA) to
large-$N_c$ QCD. In this approximation resonances are introduced
by requiring appropriate matching conditions with the Operator
Product and the Chiral Expansions. In this way a scheme
independent result is obtained.

 In the talk I reviewed the use of the MHA in
 the calculation of several electroweak matrix elements in the
Standard Model. These included the $\pi^{+}-\pi^{0}$ mass
difference\cite{Knecht:1998sp} which is a very interesting
theoretical laboratory; matrix elements of EW
penguins\cite{DeRafael:2001zs} which are of relevance in the
determination of $\epsilon'/\epsilon$\ ; the decays
$\pi^{0}\rightarrow e^{+}e^{-}$ and $\eta\rightarrow
\mu^{+}\mu^{-}$\cite{Knecht:1999gb} for which experimental
results are available; and, finally, $B_K$ in the chiral
limit\cite{Peris:2000sw} which has also been computed on the
lattice\cite{AliKhan:2001wr}.

I thank V. Cirigliano, E. Golowich and D. Lin for very interesting
discussions during the workshop and M. Golterman, T. Hambye, M.
Knecht, M. Perrottet and E. de Rafael for a very pleasant
collaboration.

\newabstract 

\begin{center}
{\large\bf Determination of $\langle (\pi\pi)_{I=2}|{\cal
Q}_{7,8}|K^0\rangle$ in the Chiral Limit}\\[0.5cm]
Vincenzo Cirigliano$^1$, John Donoghue$^2$, {\bf Gene Golowich}$^2$ 
and Kim Maltman$^3$\\[0.3cm]
$^1$Inst. f\"ur Theoretische Physik,
University of Vienna,\\
Boltzmanngasse 5, Vienna A-1090 Austria\\[0.3cm]
$^2$Physics Department, University of Massachusetts,\\
Amherst, MA 01003 USA.\\[0.3cm]
$^3$Department of Mathematics and Statistics, York University,\\
4700 Keele St., Toronto ON M3J 1P3 Canada.\\[0.3cm]
\end{center}

We discuss evaluation of the weak matrix
elements $\langle (\pi\pi)_{I=2}|{\cal Q}_{7,8}|K^0\rangle$
in the chiral limit \cite{cdgm5}. The perturbative matching is accomplished
fully within the scheme dependence used in the two loop
weak OPE calculations. The effects of dimension eight (and higher
dimension) operators are fully accounted for.  We thus obtain 
expressions for the weak matrix elements in terms of 
two dispersive sum rules which involve weights of the V-A spectral 
function.  Our numerical analysis of the sum rules is 
fortified by data from $\tau$ decay, constraints from the classical 
chiral sum rules and use of the weak OPE, and a 
careful assessment of the attendant uncertainties is given.
Our result implies that the electroweak penguin contribution 
to $\epsilon'/\epsilon$ is
\begin{equation}
\rlabel{label1}
{\epsilon' \over \epsilon}\bigg|_{\rm EWP} 
= \left( -2.2 \pm 0.7 \right) \times 10^{-3} \ \ .
\end{equation}
Comparison with other approaches is given \cite{compare}.
According to the analysis carried out in Ref.~\cite{cg},  
${\cal O}(p^2)$ corrections to our leading chiral component 
will be of order $30\%$ and negative.

\newabstract 

\begin{center}
{\large\bf Analysis of new low-energy $\pi\pi$ scattering data}\\[0.5cm]
Luca Girlanda\\[0.3cm]
Dipartimento di Fisica ``Galileo Galilei'', Universit\`a di Padova,\\
Via Marzolo 8, I35131 Padova, Italy
\end{center}

We present an analysis\cite{dfgs} of the recently published E865
data\cite{pislak} on
charged $K_{e4}$ decays and $\pi\pi$
phases, to extract values of the two s-wave scattering
lengths, of the subthreshold parameters $\alpha$ and $\beta$, of the
low-energy constants $\bar l_3$ and $\bar l_4$ as well as of the main
two-flavour order parameters: $\langle \bar u u \rangle$ and $F_{\pi}$ in
the limit $m_u=m_d=0$ taken at the physical value of the strange quark mass.
Our analysis is exclusively based on the new solutions of the Roy Equations
by Ananthanarayan, Colangelo, Gasser and Leutwyler \cite{ACGL} and on
direct experimental information.
In particular no use is made of the correlation between $a_0^0$ and $a_0^2$
as inferred from the scalar radius of the pion\cite{CGLNPB}. Instead, the
phaseshifts extracted from $K_{e4}$ data are supplemented with data on
$\delta_0^2$ below 800~MeV.
The result
\begin{equation}
a_0^0=0.228 \pm 0.012,\quad a_0^2=-0.0382\pm0.0038,
\end{equation}
is compared with the theoretical relation between $2 a_0^0 - 5 a_0^2$ and
the scalar radius of the pion obtained in two-loop standard
Chiral Perturbation Theory\cite{CGLPRL}. If the dispersive determination of
the latter, $\langle r^2 \rangle_s = ( 0.61 \pm 0.04 )$~fm$^2$ is used, one
finds a disagreement at the 1-$\sigma$ level. We argue that this discrepancy
can be explained by an unexpectedly large value of the $O(p^6)$
counterterms contributing to the above-mentioned relation. This in turn
could be a manifestation of the exceptional status of the scalar channel,
characterized by a strong $\pi\pi$ continuum and OZI rule violation.

\newabstract 

\begin{center}
{\large\bf New Approach to the Three Flavor Chiral Dynamics}\\[0.5cm]
 {\bf Jan Stern}$^1$ \\[0.3cm]
$^1$ Groupe de Physique Theorique , IPN,\\
91406 ORSAY , France\\[0.3cm]
\end{center}

As the number $N_f$ of light quark flavors increases, vacuum fluctuations of
$ \bar qq$ - pairs become more important reducing the chiral condensate and
affecting the convergence of the standard $ChPT$ expansion \cite {D}.
In practice, already two- and three- flavor dynamics may be rather
different:  In the limit $m_u = m_d = m_s = 0$ the remaining massive quarks
are heavy compared to $\Lambda_{QCD}$ and they barely affect the vacuum
structure. The corresponding three-flavor condensate
$\Sigma(3) = - <\bar qq>|_{m_u=m_d=m_s=0}$ is close to the genuine
condensate of the purely massless theory. On the other hand, the ground state
in the $N_f=2$ chiral limit $m_u=m_d=0$ with $m_s$ fixed at its physical value
($m_s \sim 150MeV \leq \Lambda_{QCD}$) is polluted by {\bf massive} $\bar ss$
pairs. The latter can modify the $SU(2)\times SU(2)$ chiral structure of the
vacuum through OZI - rule violating transition
$\bar ss \leftrightarrow\bar uu + \bar dd$ . The two-flavor condensate then
consists of two distinct pieces:
$\Sigma(2) = \Sigma(3) + m_s Z_s +\ldots > \Sigma(3)$ , where {\bf the induced
condensate} $m_s Z_s$ reflects the vacuum correlation between $0^{++}$ strange
and non-strange pairs. The latter is related to the LEC $L_6(\mu)$. Sum rule
estimates indicate that the induced condensate can be actually rather large
\cite{D}, despite its suppression in the large $N_c$ limit. We propose a
reordering of the $N_f=3$ chiral expansion based on a resummation of large
fluctuations such as $m_s Z_s$. As a first application we can study the
behavior of various order parameters (c.f
$X(N_f) = 2m\Sigma(N_f)/ M^2_{\pi} F^2_{\pi}$) in the two limiting cases:
i)For  large $N_c$ , i.e. no fluctuation , we recover the expected result
$X(3) = X(2)\sim 1$. ii) In the large fluctuation limit ( formally
realized as a large $N_f$ limit) we find $X(3)\rightarrow 0$, whereas
$X(2)$ stays close to 1 (at least for $r=m_s/m > 15$).  Numerically, the
case $N_f=3$ appears closer to the large fluctuation limit than to the large
$N_c$ limit. The non-suppression of $X(2)$ is observed in recent low-energy
$\pi\pi$ scattering experiments. We discuss how a (partial) suppression
of $X(3)$  could be seen in a combined $N_f=3$ analysis of masses, decay
constants, and of more precise $\pi\pi$ and $\pi K$ scattering data\cite{next}.

\newabstract 

\begin{center}
{\large\bf Low-energy physics of the standard model\\
  and large-$N_C$ QCD}\\[0.5cm]
Marc Knecht\\[0.3cm]
Centre de Physique Theorique,\\
CNRS Luminy Case 907, 13288 Marseille Cedex 9, France\\[0.3cm]
\end{center}

The standard model involves several different mass scales. At the very 
lower end of the spectrum, the effective degrees of
freedom reduce to the light pseudoscalar mesons, the light leptons and 
the photon. Their interactions are described by an effective lagrangian, 
whose general structure is governed by the symmetry properties of the 
standard model, but which involves scale dependent low-energy constants.
Their  values are not fixed by symmetry requirements, but refer to the 
non-perturbative QCD dynamics at the scales around $1$ GeV and above.

Starting from the study of the low-energy and high-energy behaviours
of the QCD three-point functions $\langle V\!AP\rangle$, 
$\langle VV\!P\rangle$ and $\langle AAP\rangle$, several
${\cal O}({p}^6)$ low-energy constants of the chiral Lagrangian have been
evaluated within the framework of the minimal hadronic approximation to
the large-$N_C$ limit of QCD in \cite{Knecht:2001xc}. In certain cases,
values that differ substantially from estimates based on a resonance
Lagrangian are obtained. These differences arise
through the fact that QCD short-distance constraints are in general
not correctly taken into account in the approaches using resonance
Lagrangians. Implications of our results for the ${\cal O}(p^6)$
counterterm contributions to the vector form factor of the pion and to the 
decay $\pi\to e \nu_e \gamma$, and for the pion-photon-photon transition 
form factor are discussed. The representation obtained for the latter has 
subsequently been used for a re-evaluation of the pion pole 
contribution to the anomalous magnetic moment of the muon $a_{\mu}$
(see \cite{Knecht:2001qf} and A. Nyffeler's talk). 

The hadronic light-by-light contribution to $a_{\mu}$ has also been 
discussed from the effective field theory point of view in 
\cite{Knecht:2001qg}. In particular, using a renormalization group argument, 
the coefficient of the leading logarithm arising from the two-loop 
graphs involving two anomalous WZW vertices was computed. The sign of this 
coefficient was found to be positive, in agreement with 
the evaluation of the hadronic light-by-light contribution 
to $a_{\mu}$ obtained in \cite{Knecht:2001qf}.

\newabstract 

\enlargethispage{0.5cm}
\begin{center}
{\large\bf Hadronic light-by-light scattering corrections to the muon~$g-2$:
  the pion-pole contribution}\\[0.4cm] 
Marc Knecht and {\bf Andreas Nyf\/feler}\\[0.25cm]  
Centre de Physique Th\'{e}orique, CNRS, F-13288 Marseille Cedex 9,
France\\[0.25cm]     
\end{center}

The main uncertainties in the standard model prediction for the muon
$g-2$ originate from the hadronic contributions. The hadronic
light-by-light scattering correction is particularly problematic,
since one relies purely on model calculations. According to earlier
estimates~\cite{HKS_BPP}, the dominant term involves the exchange of a
neutral pion between two anomalous vertices. Our motivation to
reevaluate~\cite{pionpole} this contribution was two-fold:
recently~\cite{resonance_estimates} we derived a new representation
for the pion-photon-photon transition form factor based on large-$N_C$
and short-distance properties of QCD. Furthermore, we wanted to push
the analytical evaluation further than in~\cite{HKS_BPP}. For a rather
general class of form factors, which also includes the VMD case and
the case of a constant form factor, derived from a point-like
Wess-Zumino-Witten vertex, we succeeded in performing all angular
integrations in the corresponding two-loop integrals analytically,
obtaining a two-dimensional integral representation for the pion-pole
contribution to $a_\mu$:
\begin{displaymath}
a_{\mu}^{\mbox{\tiny{LbyL;$\pi^0$}}} = \int_0^\infty dQ_1
\int_0^\infty dQ_2 \, \sum_i w_i(Q_1,Q_2) \, f_i(Q_1,Q_2) \, .  
\end{displaymath} 
It involves several model-independent weight-functions $w_i$, whereas
the dependence on the form factors resides in the $f_i$.  The weight
functions in the dominant term are positive definite and peaked around
$Q_1 \! \sim \! Q_2 \! \sim 0.5~\mbox{GeV}$.  This explains the
fact that $a_{\mu}^{\mbox{\tiny{LbyL;$\pi^0$}}}$ is not very sensitive
to the high-energy behavior of the form factor, but it is important to
reproduce the data at small energies.  Our result
$a_{\mu}^{\mbox{\tiny{LbyL;PS}}}= +8.3~(1.2) \times 10^{-10}$ for the
three pseudoscalar states differs essentially only by its {\it sign}
from the results given in~\cite{HKS_BPP} and has been {\it confirmed}
in the meantime~\cite{confirmation}. The new value reduces the
difference between $a_{\mu}^{\mbox{\tiny{exp}}}$ and
$a_{\mu}^{\mbox{\tiny{SM}}}$ to $1.6~\sigma$.

\newabstract 

\begin{center}
{\large\bf $N_f$-sensitivity of chiral dynamics}\\[0.5cm]
{\bf S.~Descotes$^\star$}, L.~Girlanda, N.H.~Fuchs and J.~Stern\\[0.3cm]
$^\star$ Department of Physics and Astronomy,
University of Southampton,\\
Southampton SO17 1BJ, U.K.\\[0.3cm]
\end{center}

Recent investigations in the $0^{++}$ channel~\cite{trunc} suggest 
rather different mechanisms of chiral symmetry breaking in the limits 
of $N_f=2$ massless flavours ($m_u,m_d \to 0$, but physical $m_s$) 
and $N_f=3$ ($m_u,m_d,m_s \to 0$)~\cite{nfdep}.
Meson-meson scattering is a favoured place to test 
this scenario: it is necessary to combine 
$\pi\pi$ results (2-flavour sector) including the analysis of the new $K_{e4}$ data
of the E865 collaboration~\cite{pipi}, with $\pi K$ scattering 
observables (3-flavour sector) for which dispersive estimates are now available~\cite{ABM}.

We follow the procedure outlined in~\cite{trunc} for the masses and decay constants
of $(\pi,K,\eta)$. We take their chiral expansion up to next-to-leading order in $\hat{m}$ and
$m_s$, in the isospin limit $m_u=m_d=\hat{m}$.
We can then rewrite the low-energy constants $L_{4,5,6,7,8}$ as functions of only 3 fundamental 
parameters:
$r=m_s/\hat{m}$, $X(3)=2m\Sigma(3)/(F_\pi^2M_\pi^2)$ and $F(3)=\lim_{\hat{m},m_s\to 0} F_\pi$.
We then consider the chiral expansion of $\pi\pi$ and $\pi K$ scattering observables
(namely $\alpha_{\pi\pi}$, $\beta_{\pi\pi}$ and $\beta_{\pi K}$) up to next-to-leading order, 
and reexpress $L_{4,5,6,8}$. At each stage, we keep explicitly track of the NNLO remainders 
coming from each chiral expansion.

$\pi\pi$ and $\pi K$ scattering observables constrain therefore
$[r,X(3),F(3)]$. Lines of constant $\alpha_{\pi\pi}$ and 
$\beta_{\pi\pi}$ in the $[r,X(3)]$ plane are almost perpendicular, so that
an accurate knowledge of $\pi\pi$ scattering (combined with
the $SU(3)$ chiral expansion of $F_\pi,F_K,M_\pi,M_K$) yields information about the $N_f=3$ case.
Moreover, the dispersive estimate of $\beta_{\pi K}$ favours low values of $F(3)$.
When we consider the 3 scattering 
observables altogether, nonvanishing NNLO remainders are needed for a consistent picture.
Preliminary studies favour low values of $F(3)$ (around 75 MeV) and $X(3)$ (of order 0.5
or less), and $r$ larger than 20.

\newabstract 

\begin{center}
{\large\bf Effective Chiral Lagrangian from Simulations of Partially
           Quenched QCD}\\[0.5cm]
ALPHA and UKQCD Collaborations\\[0.3cm]
Hartmut Wittig\\
DESY, Notkestr. 85, 22603 Hamburg, Germany\\[0.3cm]
\end{center}

A method to determine some of the low-energy constants in the O($p^4$)
effective chiral Lagrangian, using lattice simulations of QCD is
reviewed. It is shown how the constants $L_5$ and $L_8$ can be
determined independently by studying the quark mass dependence of
suitably defined ratios of pseudoscalar meson masses and decay
constants \cite{Heitger:2000ay}. It is pointed out that the physically
relevant coefficients can be computed even in unphysical situations,
as realised in the so-called {\em partially quenched} approximation
\citetwo{Sharpe:1997by}{Sharpe:2000bc}, as long as the physical number of
active quark flavours, $N_{\rm f}$, is employed. Furthermore, an
independent determination of $L_8$ based on the (lattice) solution of
the underlying field theory of QCD, serves to decide the question
whether or not the up-quark could be massless
\cite{mup_zero}. 

As a first step in an ultimately realistic treatment of the problem,
numerical results from simulations using $N_{\rm f}=2$ flavours are
discussed \cite{Irving:2001vy}. Results for $L_5$ and $L_8$ are
obtained with a statistical error of $5-10\%$. Systematic effects due
to neglecting higher orders are estimated to be $\pm0.25$. Future
simulations at smaller quark masses are required to corroborate this
estimate. Lattice results for $L_5$ and $L_8$ for $N_{\rm f}=2$ are
broadly consistent with the usual phenomenological determinations in
the physical three-flavour theory. These findings strongly disfavour
the possibility of massless up-quark, provided that the quark mass
dependence in the physical three-flavour case is not fundamentally
different. The method may be extended to determine $L_7$, which
involves the evaluation of disconnected contributions to mesonic
two-point functions \cite{Sharpe:2000bc}.

\newabstract 

\begin{center}
{\large\bf Radiative Corrections to $K_{\ell  3}$ Decays}\\[0.5cm]
V. Cirigliano\\[0.3cm]
Dept. of Theoretical Physics, IFIC - Univ. of Valencia,\\
P.O. BOX 22085, E-46071  Valencia, Spain\\[0.3cm]
\end{center}

In this contribution \cite{cknrt01}, I have presented general formulae
for the $K_{\ell 3}$ form factors at one loop in the framework of CHPT
with virtual photons and leptons \cite{lept}.  Two features induced by
the electromagnetic interactions are worth noticing. First, the form
factors now depend not only on the momentum transfered between the
kaon and pion, but also on a second kinematical variable. Second,
there are contributions from four of the local counterterms $X_i$,
specific to semileptonic processes \cite{lept}.

Loops with virtual photons generate also infrared divergences. In
order to deal with them, we have analysed the associated real photon
emission processes.  We have given a general description of the
changes induced in the Dalitz plot density, and have proposed a
model-independent procedure for including radiative corrections in the
data analysis.  This consists in incorporating the known long-distance
electromagnetic effects into generalized kinematical densities, while 
including all the structure dependent (UV sensitive) terms as
corrections to the form factors.  Details of the new kinematical
densities will depend eventually on the way the specific experimental
set-up deals with real photon emission.  

Within this framework, we have studied the effect of radiative
corrections in the extraction of the CKM matrix element $\vert
V_{us}\vert$ from the $K^+_{e3}$ mode.  As compared to the pure ${\cal
O}(p^4)$ form factors \cite{lr84}, the inclusion of ${\cal
O}(e^2p^2)$ electromagnetic contributions shifts $f_+(0)$ by about 
$(0.36 \pm 0.16)$\%.
Moreover, the radiative corrections produce an
effective reduction of -1.27\% for the phase space integral.  We note
here that the uncertainty on the form factor up to ${\cal
O}(p^4,e^2p^2)$ is well under control, and it affects the extraction
of $\vert V_{us}\vert$ only marginally compared to present
experimental errors. 
This result opens the road to a precision  
determination of $\vert V_{us}\vert$, for which the next 
two important ingredients are:  the inclusion of two-loop
chiral corrections and a new high statistics measurement of
branching ratios, properly accounting for radiative corrections.

\newabstract 

\newcommand{\nc}{N_{\!c}}

\begin{center}
{\large\bf Light quark mass ratios from\\
large $\nc$ chiral 
perturbation theory\footnote{Work supported by the Swiss National Science
Foundation and performed in collaboration with
H. Leutwyler}}\\[0.5cm] 
{\bf Roland Kaiser}\\[0.3cm]
Department of Physics, University of California at San Diego,\\
9500 Gilman Drive, La Jolla, CA 92093\\[0.3cm]
\end{center}

If the number of colours in QCD ($\nc$) is treated as large, the spectrum 
involves an additional light degree of freedom because the mass of the
$\eta'$ disappears in the large $\nc$ limit: $M^2_{\eta'} = O(1/\nc)$. Also in
this case, the low energy properties of QCD may be studied by means of an
effective theory, see \cite{Large N ChPT} and the references therein. To
organize the expansion, we introduce a counting parameter $\delta$ and assign
the following weights to powers of momenta, quark masses and $1/\nc$:
\begin{equation}
p = O(\sqrt{\delta}) \;\;,\;\; m = O(\delta) \;\;,\;\; 1/\nc = O(\delta) \;.
\end{equation}

Within the given scope, we performed a calculation of the mass spectrum, the
decay constants and the electromagnetic decay rates of the Goldstone particles.
The calculation is valid to next-to-leading order in the expansion in $\delta$
and also includes the logarithmically enhanced contributions of the subsequent
order \cite{in prep}.

The framework has the property that it does not allow the  
Kaplan--Manohar transformation
 \citethree{Large N ChPT}{Kaplan:1986ru}{JLab}. This
implies that the effective representations for the various observables do lead 
to constraints on the quark mass ratios when compared with the experimental
measurements. In this way, one may in particular express the ratio 
$S \equiv 2 m_s/(m_u+m_d)$ in terms of measured 
quantities and a single 
constant which determines the SU(3) breaking in the  
decay amplitudes $\eta, \eta'\rightarrow \gamma \gamma$. The low energy 
constant may be estimated theoretically by means of resonance saturation of a 
three point function \cite{Moussallam:1994xp}. 
The numerical result $S =26 \pm 2$ \cite{in prep} lies remarkably close to the
current algebra prediction $S = 25.9$.

\newabstract 

\begin{center}
{\large\bf Recalculation and Reanalysis of K $\rightarrow$ 3$\pi$.}\\[0.5cm]
Johan Bijnens, Pierre Dhonte and {\bf Fredrik Persson}\\[0.3cm]
Department of Theoretical Physics, Lund University,\\
S\"olvegatan 14A, S - 223 62 Lund, Sweden\\[0.3cm]
\end{center}
\noindent
The theoretical study of kaon decay is hindered by the nonperturbative 
behaviour of QCD in the low-energy domain. One of the approaches succesfully 
applied in the study of low-energy processes is Chiral Perturbation Theory. 
Using this, we have recalculated the CP conserving K $\rightarrow$ 3$\pi$ 
amplitude to one loop in the isospin limit. Using various symmetries
we have then figured out a way to reduce these amplitudes into publishable 
form, see \cite{Persson}. From the amplitudes we have calculated 
various decay rates and cross-sections which are compared with recent 
experimental data from the CPLEAR, HYPERON and NA48 collaborations. For this 
comparison we have used the data as they are presented in \cite{Cheshkov}.  
The result is a new fit of the $p^4$ nonleptonic weak constants in the 
Chiral Lagrangian. The $K_i$ presented in the table below are defined as 
linear combinations of these weak constants, see \cite{Kambor} and 
\cite{Ecker}. The three
columns show the preliminary results we get for three different sets of 
input data, the strong constants $L_1 - L_8$. In the first column we use the 
values presented in the 
DA$\Phi$NE book. In the second and third column the input data are taken from a
one-- and two--loop fit respectively, see \cite{Bijnens}. In the present 
results there seem to be some disagreement compared to the numbers presented 
in \cite{Kambor}.  
\begin{center}
\begin{tabular}{|l|r|r|r|r|}
\hline
Constant & $p^2$ & $p^4$ (DA$\Phi$NE) & $p^4$ (1-l. ABT) & 
$p^4$ (2-l. ABT)\\ 
\hline
$C\, G_8\, F_0^4/f_{\pi}^2$ & $8.50\times 10^{-10}$ & $4.23\times 10^{-10}$ & 
$4.23\times 10^{-10}$ & $4.23\times 10^{-11}$\\
$C\, G_{27}\, F_0^4/f_{\pi}^2$ & $4.72\times 10^{-11}$ & $3.01\times 10^{-11}$&
 $3.01\times 10^{-11}$ & $3.01\times 10^{-11}$\\
$K_2$ & - &  $5.62\times 10^{-9}$ & $6.56\times 10^{-9}$ & 
$5.55\times 10^{-9}$\\
$K_3$ & - &  $6.97\times 10^{-9}$ & $7.39\times 10^{-9}$ & 
$9.89\times 10^{-9}$\\
$K_5$ & - &  $4.34\times 10^{-12}$ & $6.30\times 10^{-13}$ & 
$4.63\times 10^{-12}$\\
$K_6$ & - &  $1.34\times 10^{-10}$ & $1.37\times 10^{-10}$ & 
$1.57\times 10^{-11}$\\
$K_7$ & - &  $-4.54\times 10^{-10}$ & $-4.55\times 10^{-10}$ & 
$-4.55\times 10^{-10}$\\
\hline
\end{tabular}
\end{center}

\newabstract 
\begin{center}
{\large\bf The $\mbox{\boldmath$\eta' \rightarrow \eta \pi \pi$}$ decay 
in $\mbox{\boldmath$U(3)$}$ chiral perturbation theory}\\[0.5cm]
N. Beisert and {\bf B. Borasoy} \\[0.3cm]
Physik Department\\
Technische Universit{\"a}t M{\"u}nchen\\
D-85747 Garching, Germany\\[0.3cm]
\end{center}

The dominant hadronic decay of the $\eta'$, 
$\eta' \rightarrow \eta \pi \pi$, is 
investigated up to one-loop order in $U(3)$ chiral perturbation theory \cite{dec}.
Within this framework, the $\eta'$ is included without employing
1/$N_c$ counting rules and loop integrals are evaluated using infrared
regularization which preserves Lorentz and chiral symmetry.
Important features of the 
$\eta$-$\eta'$ system, such as $\eta$-$\eta'$ mixing \cite{mix}, are incorporated,
and loop integrals of an $\eta'$ are shown to be suppressed.
Reasonable agreement with data is obtained without finetuning any parameters.
The investigation 
of the $\eta'$ meson may eventually lead to
a better understanding of the role of gluons in chiral dynamics due to
the axial $U(1)$ anomaly.

\newabstract 

\begin{center}
{\large\bf The Polarizability of the Nucleon:\\ The View of
Dispersion Relations}\\[0.5cm]
 Dieter Drechsel,\\[0.3cm]
 Institut f\"ur Kernphysik, Universi\"at Mainz,\\
 55099 Mainz, Germany\\[0.3cm]
\end{center}

Real and virtual Compton scattering has recently attracted much
attention on the sides of both theory and experiment. Considerable
experimental progress has resulted in improved data on the
polarizabilities from real Compton scattering (RCS), and first
data have been obtained for generalized polarizabilities from
virtual Compton scattering (VCS). In principle, polarizabilities
can be read off a low-energy expansion of the respective
scattering amplitudes in terms of small (real) photon momenta. In
order to extract precision values, however, it is useful and
mostly even necessary to analyze the data by means of dispersion
relations (DR). Therefore, we have set up both nonsubtracted and
subtracted DR for the RCS~\cite{Dre00} and VCS~\cite{Pas01}
amplitudes, thus relating these experiments with information on
the reactions $\gamma+N\rightarrow\gamma+N+X$ (s-channel
contributions) and $\gamma+\gamma\rightarrow Y$ (t-channel
contributions). The necessary input for this scheme is taken from
phase shift analyses, dispersion theory and phenomenological
models describing these reactions.

There are two immediate advantages in using DR for such analyses.
First, the kinematical range of the experiments can be extended to
momenta beyond the applicability of low-energy expansions, which
may increase the experimental sensitivity to the polarizabilities
by large factors. Second, DR allow one to evaluate higher order
polarizabilities~\cite{Hol00}, which can be predicted and thus
compared to the results from ChPT but cannot be extracted from RCS
and VCS directly. Furthermore, DR can simulate the practically
impossible experiment of doubly virtual Compton scattering,
related to quark structure functions and generalized
Gerasimov-Drell-Hearn and Burkhardt-Cottingham integrals
(Ref.~\cite{Dre01}).

\newabstract 

\begin{center}
{\large\bf Positivity Bounds for Parton
Distributions in Multicolored QCD}\\[0.5cm]
{\bf P.V. Pobylitsa}$^{1,2}$ and M.V. Polyakov$^{1,2}$\\[0.3cm]
$^1$Institute for Theoretical Physics II, Ruhr University Bochum,\\
44780 Bochum, Germany\\[0.3cm]
$^2$Petersburg Nuclear Physics Institute,
Gatchina, 188350  Russia\\[0.3cm]
\end{center}

The twist-2 parton distributions --- unpolarized $q_f(x)$, polarized $\Delta
_{L}q_f(x)$ and transversity $\Delta _{T}q_f(x)$ --- obey the well
known positivity bounds
\begin{equation}
|\Delta _{L}q_{f}|\leq q_{f}\,,\qquad 2|\Delta _{T}q_{f}|\leq q_{f}+\Delta
_{L}q_{f}\,.
\end{equation}
The second inequality is known as Soffer inequality \cite{Soffer-95}. In the
large $N_{c}$ limit the above inequalities can be enhanced as follows
\cite{Pobylitsa-2000}
\begin{equation}
|\Delta _{L}q_{u}|\le \frac13 q_{u}\,,\qquad
2|\Delta _{T}q_{u}|\leq \frac{1}{3}q_{u}+\Delta _{L}q_{u}\qquad
(N_{c}\rightarrow \infty )
\,.
\label{Soffer-enhanced}
\end{equation}
The factor of $1/3$ has nothing to do with $1/N_{c}$. The
derivation of inequalities (\ref{Soffer-enhanced}) uses no model assumptions
and is based on the
spin-flavor symmetry \citetwo{Witten-83}{Balachandran-83} due to which in
the leading order of the $1/N_{c}$ expansion the baryons with spin equal to
isospin are degenerate.

\newabstract 

\begin{center}
{\large\bf Chiral Effective Field Theory with Explicit Spin 3/2 Degrees of Freedom---A Status Report}\\[0.5cm]
Thomas R. Hemmert\\[0.3cm]
Physik Department T39, TU M{\" u}nchen\\
D-85747 Garching, Germany\\[0.3cm]
\end{center}

Non-relativistic Baryon Chpt has been very successful in describing near-threshold properties in the single Baryon sector
during the 1990s. However, when it comes to magnetic or spin-observables, the corresponding perturbative chiral expansion sometimes
behaves rather poorly (e.g. $\gamma_i$, GDH* sum-rules,\ldots) or shows rather delicate cancellations of unnaturally 
large higher order contributions (e.g. $\beta_{M1}$, $P_i^{\pi^0 p}$,\ldots). Typically,
in these channels one expects rather large contributions stemming from the first nucleon resonance, $\Delta$(1232). Therefore, in the mid 1990s
a chiral effective field theory with explicit spin 3/2 degrees of freedom {\em and systematic powercounting} was developed \cite{letter}. An 
overview of recent calculations can be found in \cite{NSTAR}. The appearance of explicit Deltas in the calculation leads to a resummation of
the chiral expansion, pushing these important physical contributions into lower orders of perturbation theory. These can result in the desired 
effects (e.g.  in the case of $\gamma_i$ \cite{spin}), but it can also upset the delicate cancellation pattern of Baryon Chpt (e.g. as in $\beta_{M1}$ \cite{beta}) . 
In the latter case we argue that additional large higher order effects should also be promoted into lower orders of perturbation theory via an 
``anomalous 
counting'' of corresponding higher order counterterms. First examples of this new development are presented 
in \cite{NSTAR} for the case of the
quark-mass dependence of the nucleons´ anomalous magntic moments and in \cite{harald} for the study of resonance contributions 
 in dynamical polarizabilities of the nucleon \cite{preprint}.

\newabstract 

\begin{center}
{\large\bf Analysis of the Pion--Kaon Sigma Term}\\[0.5cm]
Matthias Frink\\
Forschungszentrum J\"ulich\\[0.3cm]
\end{center}

To parameterize the interaction of the kaon with the non--strange scalar isoscalar source $\hat{m}(\bar{u}u+\bar{d}d)$, the scalar formfactor $\Gamma_K$ of the kaon is defined via
\begin{equation}
 \langle K_a(p_2)|\hat{m} \bigl(\bar{u}u+\bar{d}d\bigr)|K_b(p_1) \rangle=\delta_{ab} \Gamma_K(t)~,
\end{equation}
where $t=(p_1-p_2)^2\,$. The corresponding sigma term reads $\sigma_{\pi K}=\Gamma_K(0)/(2 M_K)$.
This quantity is of interest e.g.\ as a building block of the nucleon sigma term, since it is related to the strangeness content of the kaon, etc. 

A one--loop calculation in $SU(3)\,$ ChPT yields a $16 \%$ correction to the tree level result at $t=4 M_{\pi}^2$, which is fairly small as compared to the corresponding corrections to the pion sigma term in $SU(2)\,$ ChPT given in \cite{GM}.

A low--energy theorem \cite{BPP} gives the decomposition 
\begin{equation}
F^2 T^{+}_{\pi K}(s=u,\,t=2 M_{\pi}^2)=\Gamma_K(t=2M_{\pi}^2)+\Delta_{\pi K}~
\end{equation}
for the so--called Cheng--Dashen point $s=u,\,t=2 M_{\pi}^2$. $ T^{+}_{\pi K}$ is the isospin--even $\pi K$ scattering amplitude. The so--called remainder $\Delta_{\pi K}$ vanishes to leading order.
The analogue of this low--energy theorem for the $\pi \pi$ case has been analyzed in \cite{Gassain}. In an $SU(3)$ framework, the squared decay constant $F^2$ can be chosen to be $F_{\pi}^2$, $F_{\pi}F_K$, or $F_K^2$. Numerically, the first of these conventions is distinguished by a relative size of the remainder of between $1\%$ and $2\%$, as against $10\%$ to $20\%$ for the other choices. The smallness of $\Delta_{\pi K}$ for the $F_{\pi}^2$ normalization is due to a complete dropout of any terms proportional to $M_K$, which indicates $SU(2)$ breaking effects. A possible tool to separate quantities undergoing real $SU(3)$ breaking is heavy kaon ChPT as worked out in \cite{Roessl}. A further investigation is in progress.

A two--loop calculation of $\Gamma_K$ with dispersive techniques as performed for the pion case in \cite{GM} yields corrections to the tree level result of about $5\%$ at the two--pion threshold.

\newabstract 

\begin{center}
{\large\bf Soft pion theorems for partons}\\[0.5cm]
{\bf M. V. Polyakov}$^{1,2}$\\[0.3cm]
$^1$Petersburg Nuclear Physics Institute, Gatchina, 188350 Russia,\\[0.3cm]
$^2$Institute for Theoretical Physics II, Ruhr University, 44780 Bochum,Germany\\[0.3cm]
\end{center}

We show that for the
processes $\gamma^*(q) N(p)\to \pi(k) N(p')$ near the
threshold\footnote{
$M_{\pi N}-(M_N+m_\pi) \sim m_\pi$ where $M_{\pi M}$ is the invariant mass of the
produced $\pi$ and nucleon} the two
limits $Q^2\to \infty$ ($Q^2=-q^2$) and $m_\pi\to 0$ do not commute.
For $Q^2\ll \Lambda^3/m_\pi$ ($\Lambda$ is a typical hadronic scale)
one can apply classical soft pion theorems by Nambu~{\em et al.}
\cite{Nambu} which express the threshold amplitudes of the $\gamma^*N\to \pi N'$
reaction in terms of e.m. and axial form factors of the nucleon.
In the kinematics where $Q^2\gg \Lambda^3/m_\pi$ one can derive
new soft pion theorems \cite{Pobylitsa:2001cz} which also express
the threshold amplitudes in terms of the nucleon form factors.
The particular expressions depend on the form of the light-cone wave
functions (LCWF) of the nucleon, $i.e.$ sensitive to the partonic content
of the nucleon wave function.
If we assume that the dominant
component of the nucleon LCWF is symmetric we obtain, for
instance, at $ Q^2\gg \Lambda^3/m_\pi$ we have \cite{Pobylitsa:2001cz}
\begin{eqnarray}
\nonumber
A(\gamma^*p\to \pi^0 p)|_{{\rm th}}&=&
\frac{1}{3 f_\pi}
\left(\frac 5 2\ G_{Mp}(q^2)-4 \ G_{Mn}(q^2)
\right)+O\left(\frac{m_\pi}{\Lambda} \right)
\, ,
\end{eqnarray}
which should be contrasted with \cite{Nambu}
\begin{eqnarray}
\nonumber
A(\gamma^*p\to \pi^0 p)|_{{\rm th}}&=&
\frac{g_A}{ 2 f_\pi}\
G_{Mp}(q^2) +O\left(\frac{m_\pi}{\Lambda} \right)
\, ,
\end{eqnarray}
for $\Lambda^2 \ll Q^2\ll \Lambda^3/m_\pi$.
The new soft pion theorems for hard processes are in agreement
with measurements of $\gamma^*N\to \pi N'$ reaction near threshold
by E136 collaboration \cite{Bosted:1993cc}.
We also discussed soft pion theorems for other hard exclusive
processes, see details in \cite{Polyakov:1998ze}.

\newabstract 

\begin{center}
{\large\bf Chiral dynamics in meson baryon, nucleon nucleon and weak $\Lambda N
\to NN $ interactions}\\ 
E. Oset\\ 
Departamento de Fisica Teorica and IFIC\\ 
Centro Mixto Universidad de Valencia-CSIC\\
Institutos de Investigaci\'on de Paterna, Aptd. 22085, 46071 Valencia,
Spain\\ 
\end{center}

By using the techniques of the unitary extensions of $\chi PT$ \cite{oor} many
problems at intermediate energies involving the interaction of mesons and
baryons can be tackled \cite{oor}. In this talk I reported on a few recent
problems which are interconnected and which have been studied recently. On the
first hand an extension of the work of \cite{angels}, where the meson baryon
interaction in the strangeness $S=-1$ sector was studied, has been done in
\cite{cornelius}, extending the predictions of the chiral model to higher
energies. The interesting findings have been that in addition to the $\Lambda
(1405)$ resonance obtained in \cite{angels} and \cite{ulfi}, two other resonances,
 the $\Lambda (1670)$ and the $\Sigma (1520)$ were also generated dynamically 
 from the lowest order chiral Lagrangian. A remarkable finding is that the 
 $\Lambda (1670)$ resonance qualifies as a quasibound state of $K \Xi$.
 Similarly, a study has ben conducted for the $\pi N$ interaction in the region
 of the $N^*(1535)$ resonance by including the coupled channels of meson baryon
 plus also the $\pi \pi N$ channels \cite{inoue}. It was found that the phase
 shifts and inelasticities are well reproduced and that the $\pi \pi N$ channel
 was essential to reproduce the isospin $I=3/2$ sector and the $\Delta
 (1620)$ resonance.
 
   In the $N N $ interaction some progress has also been done by obtaining the
contribution of the exchange of two interacting pions in the scalar isoscalar
sector \cite{toki} using again the techniques of \cite{oor}.  We find that in
addition to a moderate attraction at intermediate distances, one also obtains a
repulsion at short distances.

 These later findings have also been used in the study of the 
 $\Lambda N \to NN $ weak transition in order to make predictions for the
 nonmesonic decay width of  $\Lambda$ hypernuclei \cite{juan} and
 particularly the ratio of neutron to proton induced $\Lambda$ decay.
 
 Finally a study of the final state interaction in the $pp \to d K^+ \bar{K^0}$
 and  $pp \to d \pi \eta$ reactions, investigated in the ANKE experiment of
 Juelich, has been performed where one finds that the  $K^+ \bar{K^0}$
 interaction as well as the $\bar{K^0} d$ interaction are very important and
 modify appreciably the invariant mass distributions \cite{ulfito}.  By fixing 
 some relevant
 parameters of the  dynamics of the process to some of the mass distributions we
 could then predict absolute rates for $\pi \eta$ production which showed
 clearly the $a_0(980)$ resonance in the $\pi \eta$
 invariant mass distribution.
\enlargethispage{0.5cm}
\begin{thebibliographynotitle}{12}
\bibitem{oor} J.A. Oller, E. Oset and A. Ramos, Prog. Part. Nucl. Phys. 45
(2000) 157
\bibitem{angels} E. Oset and A. Ramos, Nucl. Phys. A635 (1998) 99
\bibitem{cornelius} A. Ramos, E. Oset and C. Bennhold, Phys. Lett. B, in print.
\bibitem{ulfi} J.A. Oller and U.G. Meissner, Phys. Lett. B500 (2001) 263
\bibitem{inoue} T. Inoue, E. Oset and M.J. Vicente Vacas, Phys. Rev. C, in
print.
\bibitem{toki} E. Oset {\it et al.},
Prog. Theor. Phys.
103 (2000) 263.
\bibitem{juan} D. Jido, E. Oset, J.E. Palomar, Nucl. Phys. A694 (2001) 525.
\bibitem{ulfito} E. Oset, J.A. Oller and U. Meissner, Eur. Phys. J. A, in print.
\end{thebibliographynotitle}

\newabstract 

\def\si{{}^1\kern-.14em S_0}
\def\siii{{}^3\kern-.14em S_1}
\def\piii{{}^3\kern-.14em P_1}
\def\diii{{}^3\kern-.14em D_1}

\begin{center}
{\large\bf Perturbative Theories of Nuclear Forces}\\[0.5cm]
{\bf Silas R.~Beane}

Department of Physics,\\
University of Washington,\\
Seattle, WA 98195\\[0.3cm]
\end{center}

There has been recent progress in developing a consistent and converging EFT to
describe multi-nucleon systems~\cite{Be00}\cite{birarev}\cite{EKNWGM}. In very
recent work it has been shown that chiral perturbation theory in the
single-nucleon sector is a special case of a more general ordering of
operators~\cite{Be01}. In the presence of more than one nucleon, the momentum
expansion is necessarily nonperturbative at some level to accommodate the
fine-tuned scales, while the $m_q$-expansion remains perturbative. In other
words, the perturbative expansion of nuclear forces is an expansion about the
chiral limit. In the $\si$ channel only local operators are treated
nonperturbatively, whereas in the $\siii-\diii$ coupled channels it is
necessary to resum the non-local, singular part of OPE which survives in the
chiral limit.

Partial higher-order calculations in the $\siii-\diii$ coupled channels
suggest that an expansion about the chiral limit will converge. However, a full
NLO calculation is required in order to make a more definite conclusion and to
give meaningful predictions for the deuteron binding energy in the chiral
limit. The use of the square well as a short-distance regulator has proved
valuable in giving analytic formulas for the RG running of the counterterms.
However, the necessity of computing processes with external gauge fields
suggests use of a regulator that manifestly respects gauge invariance, like
dimensional regularization. An intriguing puzzle remains
concerning the relationship between square-well regularization and the matching
of delta-function interactions to singular potentials. In the short-distance limit
one finds that the singular potential wavefunctions vanish.
This makes it difficult to understand how a delta-function interaction can
modify the physical asymptotic phase.

\newabstract 

\begin{center}
{\large\bf Renormalizing the Lippmann-Schwinger equation for the OPEP}\\[0.5cm]
Dolors Eiras and {\bf Joan Soto} \\[0.3cm]
Dept. d'ECM, Universitat de Barcelona,\\
Diagonal 647, 08028 Barcelona, Catalonia, Spain\\[0.3cm]

\end{center}

Since the seminal papers of Weinberg \cite{Weinberg}, ten years have passed and there is still no consensus 
on how to organize an EFT for low energy nucleon-nucleon interactions.
We follow the general ideas of \cite{Mont} and consider the two nucleon system interacting through a 
potential, namely at energies much below the pion mass (but allowing three-momenta to be similar to it), as an EFT on its own. Although the potential 
can be derived 
order by order in $\chi$PT from the Heavy Baryon Chiral Lagrangian, 
the calculations in this EFT need not be carried out order by order in  $\chi$PT anymore. In fact the results of \cite{Stewart} 
seem to indicate, particularly for the triplet channel, that the OPEP (or at least part of it) should be taken into account at all orders of $\chi$PT.
If so, the question arises whether the renormalization program, which is usually carried out perturbatively, goes through. We have addressed this question for the OPEP in \cite{nn} and obtained the following results:

$\bullet$ The singlet channel is renormalizable.

$\bullet$ The triplet channel is non-renormalizable, unless the coupling constant of a non-local potential is allowed to flow.
          Even in that case, the renormalized amplitude has undesirable properties (lacks $^3S_1$-$^3D_1$ mixing).

$\bullet$ If a suitable piece of the potential in the triplet channel is treated as a perturbation, then it becomes renormalizable 
          at least up to next to leading order.

These results led us to propose a new way to organize the calculation in the triplet channel which, we  conjecture, will be renormalizable at any order.
We expect the new renormalized perturbative expansion to have better convergence properties than previous proposals as well \cite{nn2}.

\newabstract 

\begin{center}
  {\large\bf Dispersive Effects in Nucleon Polarisabilities}\\[0.5cm]
  Harald W.~Grie{\ss}hammer$^{1}
  $\\[0.3cm]
  $^1$Institut f{\"u}r Theoretische Physik (T39), TU M{\"u}nchen,
  Germany\\[0.3cm]
\end{center}
A formalism to extract the dynamical nucleon polarisabilities defined via a
multipole expansion of the structure amplitudes in nucleon Compton scattering
was developed in \cite{hgth}. In contradistinction to the static
polarisabilities, dynamical polarisabilities gauge the response of the
internal degrees of freedom of a composed object to an external, real photon
field of arbitrary energy. Being energy dependent, they contain information
about dispersive effects induced by internal relaxation mechanisms, baryonic
resonances and meson production thresholds of the nucleon.  Explicit formulae
to extract the dynamical electric and magnetic dipole as well as quadrupole
polarisabilities from low energy nucleon Compton scattering are presented, and
the connection to the definition of static nucleon polarisabilities is
discussed.
A chiral analysis using iso-scalar dynamical polarisabilities including the
$\Delta(1232)$ as dynamical degree of freedom is performed in~\cite{hgthrh}.
The energy dependence of $\alpha_{E1}(\omega)$ and $\beta_{M1}(\omega)$ is
well predicted by chiral dynamics. Polarisabilities stem predominantly from
charge displacements of the pion cloud around the nucleon and from the
$\Delta$ resonance contribution. However, the counter terms for
$\alpha_{E1}(\omega)$ and $\beta_{M1}(\omega)$ turn out to be at least 4 times
larger than na\"{\i}ve dimensional analysis predicts. By consistently
modifying the SSE power counting~\cite{th}, they are thus promoted to LO, and
their size is fixed by fitting to the static polarisabilities measured.  Their
energy dependence is then a prediction which is in good agreement with a
dispersion relation (DR) analysis. $\alpha_{E2}(\omega)$ is well predicted
without free parameters up to $150\;\mathrm{MeV}$, where DR shows the $E2$
$N\to\Delta$ transition to be important.  $\beta_{M2}(\omega)$ lacks overall
strength, and the DR curve shows a strong, nearly linear energy dependence
which might originate from a strong dia-magnetic quadrupole relaxation.

In Compton scattering on the deuteron at $\omega=91\;\mathrm{MeV}$, dynamical
effects are large and cannot be mimicked by taking only the slopes of the
polarisabilities at zero energy into account.  The dynamical polarisabilities
constructed above predict values in agreement with experiment, in
contradistinction to traditional analyses where $\beta_{M1}(0)$ is five times
bigger than from extractions at zero energy.

\newabstract 

\begin{center}
{\large\bf PION ELECTROPRODUCTION ON THE DEUTERON NEAR THE TRESHOLD}\\[0.5cm]
 {\bf Hermann Krebs}\\[0.3cm]
Forschungszentrum J\"ulich,\\
52425 J\"ulich, Germany.\\[0.3cm]
\end{center}

In the last years there was a lot of progress in describing low energy
physics with the help of chiral perturbation theory. One of them is the
description of $\pi^0$-electroproduction on a proton.
The absence of a neutron target makes the comparison between theory
and experiment more complicated: One has to describe the production process
 on the lightest nucleus, where neutrons are bound. The first calculation of 
$\pi^0$-electroproduction on a deuteron  within the 
framework of HBCHPT  was done at $\pi^0$-threshold in the
static limit\cite{thresh}. For comparison with experiment the data were extrapolated to
the threshold\cite{MAMI}. One obtained no agreement for the longitudinal part. 
To improve the situation I transformed the single-nucleon contribution into 
the right frame. Secondly to avoid any extrapolation of the data I calculated 
the amplitudes above the $\pi^0$-threshold, where the data were really measured. 
The calculation consists of two steps.   

The first step is the multipole decomposition. Arenh\"ovel\cite{AREN}
constructed 13 linearly independent spin structures and described the amplitude
by 13 structure functions. On the other side, because of the selection rules
there are 13 multipole amplitudes for each multipolarity, which do not depend
on the angle between the outgoing pion and deuteron. The structure function
vector $F=(F_1,\ldots,F_{13})$ can be described in the following way:
\begin{equation}
\rlabel{label1}
\begin{array}{lll}
F = \sum_{L=0}^\infty A_L(\cos(\theta))M_L & {\rm and} & M_L = \int_0^1 d \cos(\theta) B_L(\cos(\theta))F,\\
\end{array}
\end{equation}
where $M_L = (M^1_L,\ldots,M^{13}_L)$ is the multipole vector and $A_L,B_L$
are matrices that only depend on $\cos(\theta)$.

The second step is the calculation of the amplitude. For this purpose
I use the hybrid approach first introduced by Weinberg: 
one calculates the irreducible kernel within the framework of HBCHPT 
to the order $q^3$ and convolutes it with phenomenological or chiral wave-functions.
 To third order in small momenta, the amplitude is finite and 
a sum of two- and three-body interactions with no undetermined parameters.

\newabstract 
\begin{center}
{\large\bf One Two Three ... Infinity: Nucleons in EFT}\\[0.5cm]
{\bf U. van Kolck}$^{1,2}$\\[0.3cm]
$^1$Department of Physics, University of Arizona,\\
Tucson, AZ 85721, USA\\[0.3cm]
$^2$RIKEN-BNL Research Center, Brookhaven National Laboratory,\\
Upton, NY 11973, USA\\[0.3cm]
\end{center}

Effective Field Theories (EFTs) provide the only 
existing framework to derive nuclear
structure from QCD.
Because the typical momenta of nucleons in the deuteron
and in the virtual bound state in the $^1S_0$ $NN$ channel
are smaller than the pion mass, an EFT involving nucleons
only (where pions and all other mesons have been integrated
in favor of contact interactions) has been developed.
This EFT was applied to systems
with two \cite{2bod} and even three \cite{3bod} nucleons,
with much success.
An exploratory study of infinite nuclear matter
at finite temperature was also conducted with a lattice regulator
\cite{inftybod}.
 
In order to bridge the extremes in baryon number, we are now studying
halo nuclei. A lower-energy EFT
involving contact interactions among 
nucleons and a core can be formulated \cite{halos}.
One simple application is $N$-$^4$He scattering, which is known
to display a low-lying
$P_{3/2}$ resonance. 
The power counting in the case of a shallow $P$-wave state
is similar
to that \cite{2bod} for $S$-wave states relevant in $NN$ scattering,
but has some important differences.
In leading order it involves two operators, which can accommodate
resonance behavior in addition to a bound state.
One indeed finds good agreement with phase-shift analyses.
With parameters fitted this way, 
$^6$He can then be studied as a three-body system
using techniques like those of Ref. \cite{3bod}. 
Generalization
to other halo nuclei is straightforward, once cores with spin are considered.

\newabstract 

\begin{center}
{\large\bf Few-Body Physics in Effective Field Theory}\\[0.5cm]
{H.-W. Hammer}\\[0.3cm]
Department of Physics,
 The Ohio State University,\\
 Columbus, OH 43210, USA\\[0.3cm]
\end{center}

Effective Field Theory (EFT) provides a powerful framework that
exploits a separation of scales in physical systems to perform systematic, 
model-independent calculations.  The long-distance physics is included
explicitly, while the corrections from short-distance physics
are calculated in an expansion in the ratio of short- and long-distance
scales. We apply EFT to three-body systems with short-range interactions
and large two-body scattering lengths $a \gg R$,
where $R$ is the typical range of the interaction. 
These systems are interesting because they display universal features 
such as a logarithmic spectrum of shallow three-body bound 
states and a discrete scale invariance \cite{Efimov}. 
While most channels are straightforward, in certain $S$-wave channels 
(e.g. for spinless bosons) a nonderivative, one-parameter three-body force 
is required at leading order for consistent renormalization. 
The renormalization group evolution of this  
three-body force is  governed by a limit cycle  \cite{3body}. 

In nuclear physics, this EFT has successfully been applied to the 
neutron-deuteron and $\Lambda$-deuteron systems \cite{triton}. 
The variation of the three-body force provides a compelling explanation 
of the Phillips line.

Especially interesting is the application of this formalism to
the physics of cold atoms and Bose-Einstein condensates, where
the scattering length $a$ can be tuned experimentally 
using Feshbach resonances. As a consequence,
the unique $a$-dependence predicted by the EFT can be tested.
In particular, the EFT predicts logarithmically spaced
minima (for $a>0$) and resonances (for $a< 0$)
in the rate for three-body recombination, i.e. when 
two atoms form a molecule and a third atom balances energy 
and momentum \cite{bec}.

\newabstract 

\begin{center}
{\large\bf Application of Chiral Nuclear Forces to Few-Nucleon
Systems }\\[0.5cm]
{\bf W. Gl\"ockle}\\[0.3cm]
Institut f\"ur theoretische Physik II, Bochum\\
Ruhr-Universit\"at Bochum\\

\end{center}

In recent years high precision NN force models have been developed which 
are related to the OBE picture but modified by purely phenomenological
 structures. Based on typically 40 parameters they describe the rich set
 of NN data up to the pion threshold perfectly well. They alone lead to
 underbinding of nuclear ground states and 
are insufficient to describe 3N scattering observables well at energies 
starting around 60 MeV and higher. On the  other hand at low energies
 they work rather well for 3N scattering observables with the 
outstanding exception of the analysing power $Ay$ in elastic Nd scattering.
 Three-nucleon forces are natural candidates to possibly cure these
 failures. There are various models which, however, are not
 theoretically linked to the
NN potential models. They can be tuned to the $^3H$ binding energy, 
but they neither cure the $A_y$-puzzle nor remove all discrepancies in 
3N scattering at the higher energies mentioned. Clearly a more systematic
 and theoretically founded approach for nuclear forces is needed. Chiral
 perturbation theory appears to be a promising step in that direction. Based
 on an effective chiral Lagrangian for pion 
and nucleon fields one can eliminate in the spirit of the Weinberg power
 counting the pionic degrees of
freedom leading to nuclear force ordered in powers of generic momenta over 
a mass scale of the order of the $\rho$-mass. We applied these chiral
 forces in the LO, NLO
and NNLO's. Up to these orders they are formed out of one - and two-pion
 exchanges and a string of contact forces with altogether 9 LEC's. At NNLO
 the
 resulting predictions based on rigorous Faddeev-Yakubovsky calculations 
are very similar to the ones of
 the above mentioned potential models. In the chiral scenario the momenta
 are cut off (smoothly) between $\wedge = 500$ and 600 MeV/c and the
 cut-off dependence shrinks significantly in going from NLO to NNLO.
 At NNLO the first time three-nucleon forces
of three different topologies occur in that systematic approach.
 They depend on two additional parameters, which allow of course to guarantee 
the correct $^3H$ binding energy. Predictions for the $^4H$ binding energy
 and the 
wealth of 3N scattering observables is in progress. It is further planned to
 go to NNNLO and to incorporate also electroweak probes, which would
 significantly improve the theoretical foundation for electroweak many-body
 currents. There is good reason to 
hope that the chiral approach
 to nuclear forces will put nuclear physics on a theoretically better
 founded basis than hitherto.
\vskip1truecm
{\bf Acknowledgement:} This work is carried through in collaboration with E
. Epelbaum, U.G. Mei{\ss}ner, A. Nogga, H. Kamada and H. Witala.     

\newabstract 

\begin{center}
{\large\bf Nuclear Forces from Chiral EFT: Recent Developments}\\[0.5cm]
{\bf Evgeny Epelbaum}$^1$, Walter Gl\"ockle$^1$ and Ulf-G. Mei\ss ner$^2$\\[0.3cm]
$^1$Ruhr--Universit\"at Bochum, Institut f\"ur Theoretische Physik II,\\
D-44870 Bochum, Germany\\[0.3cm]
$^2$Forschungszentrum J\"ulich, Institut f\"ur Kernphysik (Theorie),\\
D-52425 J\"ulich, Germany\\[0.3cm]
\end{center}
Few--nucleon forces at various leading orders in chiral effective field theory derived 
from the most general chiral invariant Lagrangian for nucleons and pions using
the method of unitary transformation \cite{EGM1} are briefly reviewed.
To leading order, the NN potential consists of the undisputed one--pion exchange (OPE) 
and two NN contact interactions. Corrections at NLO stem from the leading chiral TPE and additional 
contact terms with two derivatives. At NNLO one has to include the subleading TPE which 
contains $\pi \pi$NN interactions with two derivatives (pion mass insertions). The large
values of the corresponding coupling constants $c_{1,3,4}$ obtained in the $Q^2$ and $Q^3$
analysis of $\pi$N scattering lead at this order 
to unrealistically strong attraction and, as a consequence, to 
unphysical deeply bound states \cite{EGM2}. We provide arguments for
a reduction of the TPE potential and introduce the NNLO* 
version of the NN forces, which is very well suited for few--nucleon calculations.
We show that the NN phase shifts are strongly improved at NNLO* compared to NLO and are mostly 
rather well reproduced up to $E_{\rm lab}=$200 MeV. 

We also establish a link between the NN potential derived in chiral effective field theory and 
various modern high--precision potentials \cite{EGME}. To be more specific, we show that  the values
of the NN contact interactions can be understood from the resonance saturation in terms of well 
established one--boson exchange models. We also address the issue of the naturalness of the contact
terms and consider leading isospin violating effects \cite{WME}.

\newabstract 
\begin{center}
{\large\bf Chiral dynamics of nuclear matter}\\[0.5cm]
Norbert Kaiser\\[0.3cm]
Physik-Department T39, TU M\"unchen, D-85747 Garching, 
Germany.\\[0.3cm]
\end{center}

We calculate the equation of state of isospin-symmetric nuclear matter in the 
3-loop approximation of chiral perturbation theory \cite{pap1}. The 
contributions to the energy per particle $\bar E(k_f)$ from $1\pi$- and
$2\pi$-exchange graphs are ordered in powers of the Fermi momentum $k_f$
(modulo functions of $k_f/m_\pi$). It is demonstrated that, already at order 
${\cal O}(k_f^4)$, $2\pi$-exchange produces realistic nuclear binding. The 
underlying saturation mechanism is surprisingly simple (in the chiral limit), 
namely the combination of an attractive $k_f^3$-term and a repulsive 
$k_f^4$-term. The empirical saturation point $\bar E(k_{f0})\simeq -15.3$\,MeV,
$\rho_0 \simeq 0.17\,$fm$^{-3}$ and the nuclear compressibility $K\simeq
255\,$MeV are well reproduced at order ${\cal O}(k_f^5)$ with a momentum
cut-off of $\Lambda =7f_\pi \simeq 0.65$\,GeV which parametrizes all necessary
short-range dynamics. Next, we calculate the density-dependent asymmetry energy
and find $A(k_{f0})\simeq 34\,$MeV, in very good agreement with the empirical 
value. The pure neutron matter equation of state is also in fair agreement 
with sophisticated many-body calculations and a resummation result of effective
field theory, for low neutron densities $\rho_n <0.25\,$fm$^{-3}$.  

In the same framework we evaluate the momentum and density dependent complex 
single particle potential $U(p,k_f)+i\,W(p,k_f)$ \cite{pap2}. $1\pi$- and 
$2\pi$-exchange give rise to a potential depth (for a nucleon at the bottom of 
the Fermi sea) of $U(0,k_{f0}) = -53.2\,$MeV, in agreement with the shell model
potential. The momentum dependence of the real part $U(p,k_{f0})$ is
non-monotonic and can be translated into a mean  effective nucleon mass of 
$\bar M^* \simeq 0.82 M$. The imaginary part $W(p,k_f)$ is generated entirely
by iterated $1\pi$-exchange. The half width of a nucleon hole with $p=0$ comes
out as $W(0,k_{f0})=29.7\,$MeV. The basic theorems of Hugenholtz-Van-Hove and
Luttinger are satisfied in our calculation. 

We also consider nuclear matter finite temperature $T$. The 
free energy per particle $\bar F(\rho,T)$ is obtained from the energy density 
functional by inserting a Fermi-Dirac distribution for the density of states. 
The calculated pressure isotherms indicate a liquid-gas phase 
transition at $T_c \simeq 26\,$MeV and $\rho_c \simeq 0.09\,$fm$^{-3}$. 

Finally, we calculate the nuclear spin-orbit strength from the spin-dependent 
part of the interaction energy $\Sigma_{spin} = {i\over 2}\, \vec \sigma \cdot
(\vec q \times\vec p\,) \, U_{ls}(p,k_f)$ of a nucleon scattering off weakly 
inhomogeneous nuclear matter. We find from iterated $1\pi$-exchange $U_{ls}(0,
k_{f0})=  35.1$\,MeVfm$^2$ in perfect agreement with the shell model value.

\newabstract 

\begin{center}
{\large\bf Chiral Perturbation Theory at Finite Density. Beyond the 
Mean Field Approach}\\[0.5cm]
Ulf-G. Mei\ss ner$^1$, {\bf Jos\'e A. Oller}$^2$ and Andreas Wirzba$^1$\\[0.3cm]
$^1$ Forschungszentrum J\"ulich, Institut f\"ur Kernphysik\\
D-52425 J\"ulich, Germany\\[0.3cm]
$^2$ Departamento de F\'{\i}sica, Universidad de Murcia\\
E-30071 Murcia, Spain
\end{center}

An explicit expression for the generating functional of two--flavor
    low--energy QCD with external sources in the presence of non-vanishing
    nucleon densities has been derived recently~\cite{med1}.  Within this
    approach we derive \cite{med2} power counting rules for the calculation 
    of in-medium
    pion properties.  We develop the so-called standard rules for residual
    nucleon energies of the order of the pion mass and a modified scheme
    (non-standard counting) for vanishing residual nucleon energies. We also
    establish the different scales for the range of applicability of this
    perturbative expansion, which are $\sqrt{6}\pi f_\pi\simeq 0.7$ GeV for
    the standard and $6\pi^2 f_\pi^2/2m_N\simeq 0.27$ GeV for non-standard
    counting, respectively.  We have performed a systematic analysis of
    n--point in-medium Green functions up to and including next-to-leading
    order when the standard rules apply. These include the in-medium
    contributions to quark condensates, pion propagators, pion masses and
    couplings of the axial-vector, vector and pseudoscalar currents to pions.
    In particular, we find a mass shift for negatively charged pions in heavy
    nuclei, $\Delta M_{\pi^-}=(18\pm 5)\,{\rm MeV}$, 
    that agrees with recent determinations from deeply bound pionic
    $^{207}$Pb.  We also show that  the unique 
    role of $f_\pi$  as order parameter of chiral symmetry 
    breaking in vacuum corresponds to $f_t$ in symmetric nuclear matter. The latter is the coupling of 
    a pion at rest to the temporal component of the axial-vector current. In 
    addition, we have  established the absence of in-medium
    renormalization in the $\pi^0 \to \gamma\gamma$ decay amplitude up to the
    same order. The study of $\pi\pi$ scattering requires the use of the
    non-standard counting and the calculation is done at leading order. Even
    at that order we establish new contributions not considered so far.  We
    also point towards further possible improvements of this scheme and touch
    upon its relation to more conventional many-body approaches.

\newabstract 

\begin{center}
{\large\bf Color superconductivity in high-density QCD}\\[0.5cm]
Mark Alford \\[0.3cm]
Physics and Astronomy Dept, Glasgow University, Glasgow G12 8QQ, UK.
\\[0.3cm]
\end{center}

At high quark density, it is expected that a condensate
of Cooper pairs of quarks will form, 
spontaneously breaking the
$SU(3)$ color gauge symmetry. This is known as color
superconductivity. Since the QCD interaction between quarks
is strongly attractive, the pairing is expected to be characterized 
by a large gap, in the range of tens to hundreds of MeV.
For recent reviews, see the introduction to this miniproceedings.

In matter above nuclear density, the quark chemical potential
$\mu \gtrsim 400~{\rm MeV}$, which is greater than the ``bare''
mass of the strange quark but less than that of the charm quark, so it
is reasonable to consider 3 flavor quark matter. This leads to a 
``color-flavor locked'' (CFL) pattern of quark pairing \cite{Alford:1999mk}, which
has many interesting properties. It breaks chiral symmetry
without a $\langle \bar q q\rangle$ condensate,
causes the photon to mix with one of the gluons, and
may be continuously connected to lower density hyperonic matter
(``quark hadron continuity'').

A compact star is the best place to look for CFL quark matter in nature,
since it is dense and cold. Recently there has been work on the
expected structure of such a star \cite{Alford:2001zr}.
The two possibilities are (a) a sharp interface between a CFL core
and a nucler matter mantle, and (b) a layer of mixed phase,
consisting of positively charged nuclear matter mixed with negatively
charged CFL quark matter. In Ref.~\cite{Alford:2001zr} we found that
the sharp interface would be favored if the surface tension of the
nuclear-CFL interface $\sigma \gtrsim 40~{\rm MeV/fm}$.

The next step is to find observable phenomena that indicate the
internal structure of the star. The sharp interface creates a
discontinuity in the density by a factor of two, so as well as affecting
the mass-radius relation it may lead to signatures in the gravitational waves
emitted in the merger of two such stars, which are supposed to be
detectable by future gravitational wave detectors such as LIGO II.
The mixed phase has structure on the scale of a few fm, 
which is expected to reduce the mean free path of neutrinos,
leading to possible signatures in the time profile
of neutrinos emitted during the supernova that 
creates the compact star.  Work on these topics is in progress.

\newabstract 

\begin{center}
{\large\bf Crystalline Color Superconductivity}\\[0.5cm]
Krishna Rajagopal\\[0.3cm]
Center for Theoretical Physics, MIT\\
Cambridge, MA, USA 02139\\[0.3cm]
\end{center}

At asymptotic quark number densities, the ground state of QCD
is expected to be the color-flavor locked (CFL) 
phase, in which all quarks participate in BCS pairing~\cite{CFL}.
Below some critical density determined by quark mass differences,
CFL is disrupted, leaving some species of quarks with 
differing Fermi momenta unpaired.
It is natural to ask whether there is some generalization
of the BCS ansatz in which pairing 
between two species of quarks persists even once their
Fermi momenta differ.   Crystalline color superconductivity
is the answer to this question.
The idea is that it may 
be favorable for quarks with differing Fermi momenta 
to form pairs whose momenta are {\it not} equal
in magnitude and opposite in sign~\citetwo{LOFF}{BowersLOFF}.  
Condensates of this sort spontaneously
break translational and rotational invariance, leading to
gaps which vary in a crystalline pattern.
The favored crystal structure can be determined by fixing the
coefficients in a 
Ginzburg-Landau EFT; this is the subject of ongoing work.
The phonons of this phase can be analyzed using EFT 
methods~\cite{LOFFphonon}.
The window in parameter space (or in density)
in which crystalline color superconductivity arises is
narrow in toy models, but this phase is generic in QCD~\cite{pertloff}.
If in some shell within the quark matter core
of a neutron star 
the quark number densities are
such that crystalline color superconductivity arises,
rotational vortices may be pinned in this shell, making
it a locus for glitch phenomena.

\newabstract 

\begin{center}
{\large\bf The Prefactor of the Color-Superconducting Gap}\\[0.5cm]
Dirk H.\ Rischke\\[0.3cm]
Institut f\"ur Theoretische Physik, J.W.\ Goethe--Universit\"at,\\
Robert-Mayer-Str.\ 10, D--60054 Frankfurt am Main, Germany\\[0.3cm]
\end{center}

The magnitude of the color-superconducting gap $\phi$ in cold, dense quark
matter is computed via a gap equation. At the Fermi surface and
for $g \ll 1$ ($g$ is the QCD coupling constant),
\begin{equation}
\phi = b\, \mu \, \exp \left( - \frac{c}{g} \right)\,
\left[ 1 + O(g) \right]\,\,,
\end{equation}
where $\mu$ is the quark chemical potential, $c = 3 \pi^2/\sqrt{2},$ and
$b = 512 \pi^4\, [ 2/(N_f\,g^2)]^{5/2}$ $\times \exp 
\left[ - (\pi^2 + 4)/8 \right]$, $N_f$ is the number of
degenerate quark flavors.
The constant $c$ is determined by almost static magnetic
gluon exchange \cite{son}, which is a long-range interaction in QCD.
The constant $b$ is determined by static electric and non-static
magnetic gluon exchange \cite{rdpdhr}, as well as by contributions from the
quark self-energy \citetwo{rock}{qwdhr}.
$O(g)$ corrections to the constant $b$ arise for instance 
from non-static electric gluons \cite{rdpdhr}, vertex
corrections \cite{rock}, the color-Meissner effect \cite{dhr}, 
and from the finite lifetime of
quasi-particles off the Fermi surface \cite{manuel}.
At this order, there exists also an apparent gauge dependence of
the gap equation in mean-field approximation \cite{gauge}.
The effect of the running of the coupling constant \cite{seattle}
on $b$ is unclear at present.

\newabstract 

\begin{center}
{\large\bf Effective Theories for QCD at high density}\\[0.5cm]
Roberto Casalbuoni\\[0.3cm]
Dep. Theor. Phys. 2, University of Florence\\
I-50019 Firenze, Italia\\[0.3cm]
\end{center}
The effective lagrangian  description \cite{1} of two phases of
QCD at high density is  presented. We  describe the CFL phase,
realized at very high density in the case of three massless
quarks, and the LOFF phase, a possible crystalline state of QCD
matter at intermediate density. In both cases we construct the
relevant low-energy effective lagrangian and show how to evaluate
perturbatively the couplings. These calculations have been
performed by using the idea of gapped quasi-fermionic excitations
close to the Fermi surface \cite{2}. We couple these excitations
to the Nambu Goldstone bosons (NGB) and evaluate the corresponding
relevant Green functions, obtaining  the couplings. The main
advantage of this description is in the fact that close to the
Fermi surface the physics reduces to infinite copies of a
two-dimensional problem. For the CFL phase we evaluate the NGB
decay constants and their velocity (being at finite density,
Lorentz invariance is broken). We also evaluate the gluon
self-energy showing that the gluon physical masses are not of
order $g_s\mu$ ($\mu$ being the chemical potential) but rather of
the order of the gap $\Delta$. This effect is due to a very large
wave function renormalization effect of order $g_s\mu/\Delta$.
Finally we discuss the LOFF phase. In this case translational and
rotational invariance are broken, but we show that only one
physical NGB appears, the phonon. On the basis of the effective
lagrangian approach, the phonon shows a peculiar anisotropic
dispersion relation. Work in progress \cite{3} about the
evaluation of the anisotropy is also presented.

\newabstract 

\begin{center}
{\large\bf Dense QCD}\\[0.5cm]
{\bf Ismail Zahed}\\[0.3cm]
Department of Physics and Astronomy,\\
State University of New-York, Stony Brook NY 11794.\\[0.3cm]
\end{center}
QCD at asymptotic densities exhibits a color superconducting state with
color-flavor locking (CFL) and parametrically large gaps thanks to the
long-range character of the colormagnetic interaction~\cite{CFL}. 
At non-asymptotic densities, QCD may crystallize in either an insulating
or a superconducting phase~\cite{OVERLOFF}. The CFL phase supports light
Goldstone modes in the form of particle-particle and hole-hole
excitations with anomalous radiative decays. Very dense QCD
is ideally suited for an effective  Lagrangian approach with coefficient
constants calculable from first principles~\cite{HONGWIR}. One way to
investigate the physical nature of this phase as well as the general
character of the phase diagram in QCD is to quantify the amount of
electromagnetic emissivities. Although these emissivities are strictly
calculable at asymptotic densities, it is always intriguing to
extrapolate them at few times that of normal nuclear matter. For large
superconducting gaps as suggested by both perturbative and
nonperturbative estimates, the dilepton and photon rates from the CFL
phase at temperatures of the order of $T=80$ MeV are comparable to the
one extrapolated from the low density hadronic phase~\cite{LEPTON}.

\newabstract 

\enlargethispage{0.5cm}
\begin{center}
{\large\bf Color Superfluidity in Two Color QCD on the Lattice}\\[0.5cm]
E. Bittner$^{1}$, M.-P. Lombardo$^{2}$, {\bf H. Markum}$^1$ and R. Pullirsch$^{1}$\\[0.3cm]
$^1$ Atominstitut der \"osterreichischen Universit\"aten, Techn. Univ.  Wien,\\
A-1040 Vienna, Austria\\[0.3cm]
$^{2}$ Istituto Nazionale di Fisica Nucleare, Sezione di Padova,\\
and Gruppo Collegato di Trento, Italy\\[0.3cm]
\end{center}

We analyze the eigenvalue spectrum of the staggered Dirac matrix
in two-color QCD at nonzero chemical potential $\mu$ when the
eigenvalues become complex. The quasi-zero modes and their role for
chiral symmetry breaking and the deconfinement transition are examined, see Figure 1.
The density $\rho(y)$ is used to estimate a value for the chiral condensate by
applying the Banks-Casher relation (which originally was derived for real
eigenvalues appearing in pairs of opposite sign). We further employed the
standard definition of the Green function by inverting the fermionic matrix
with a noisy source and by computing its eigenvalues exactly, respectively,
to get the condensate.
An analogous analysis is performed for the spectrum of the Gor'kov
representation of the fermionic action yielding the diquark condensate~[1].

\noindent
[1] E.~Bittner, M.-P.~Lombardo, H.~Markum, R.~Pullirsch,
Nucl. Phys. B (Proc. Suppl.) 94 (2001) 445


\begin{figure*}[h]
   \begin{center}
     \epsfig{figure=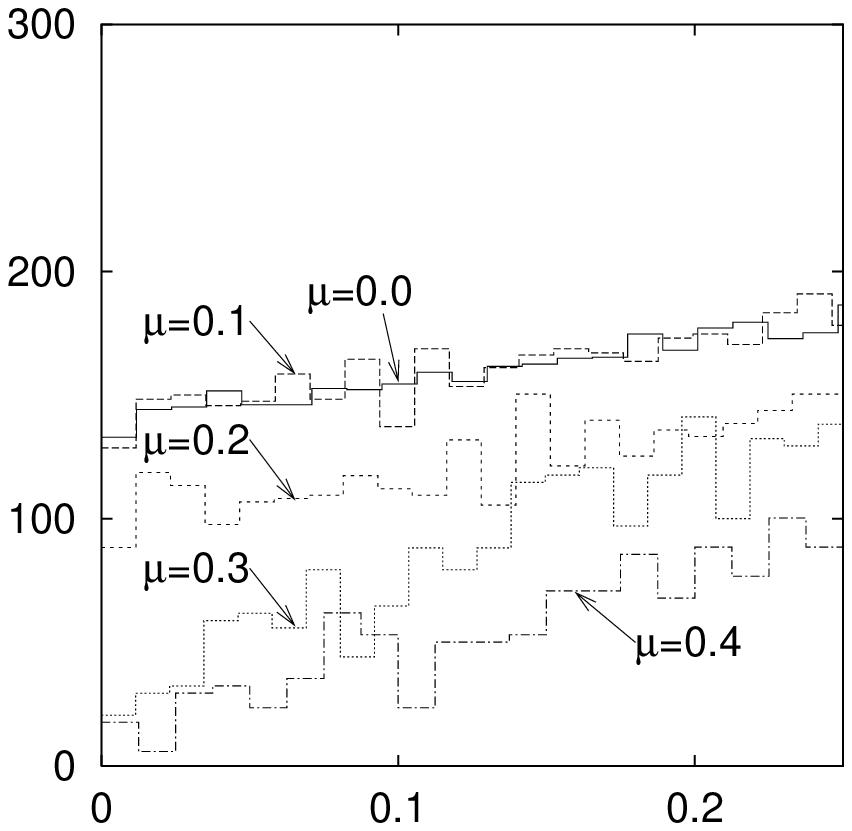,width=4.9cm,height=3.8cm}\hspace*{15mm}
     \epsfig{figure=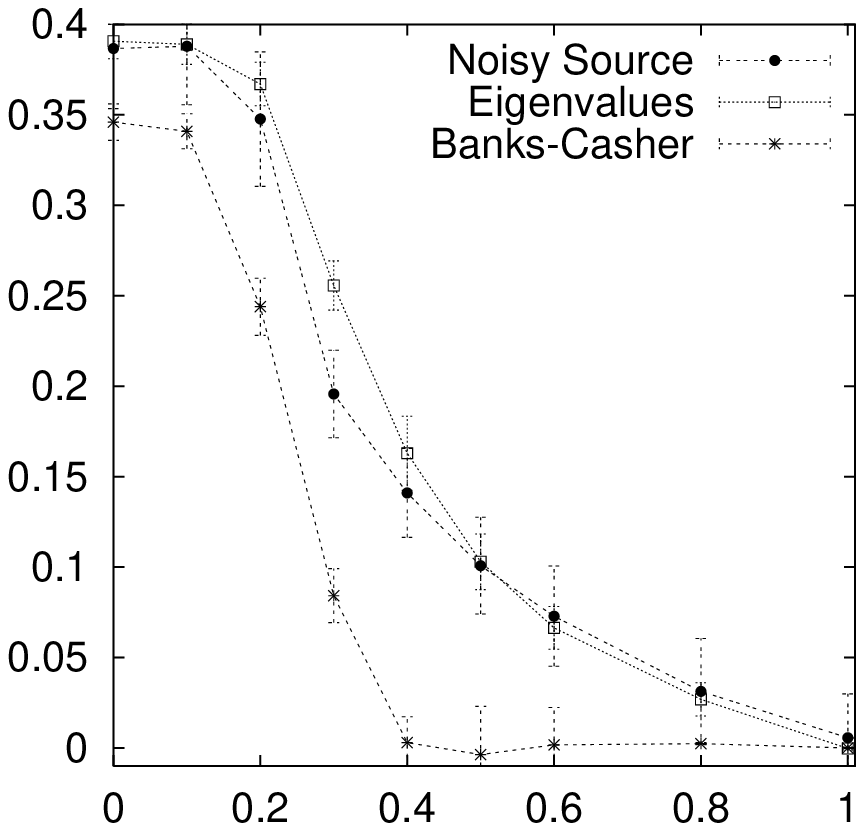,width=4.9cm,height=3.8cm}\\[-42mm]
     \hspace*{-50mm}\phantom{$\rho(y)$}\hspace*{50mm}\phantom{$\langle\bar\psi\psi\rangle$}\\[0.5mm]
     \hspace*{-54mm}$\rho(y)$\hspace*{59mm}$\langle\bar\psi\psi\rangle$\\[29mm]
   \end{center}
   \begin{center}
     \epsfig{figure=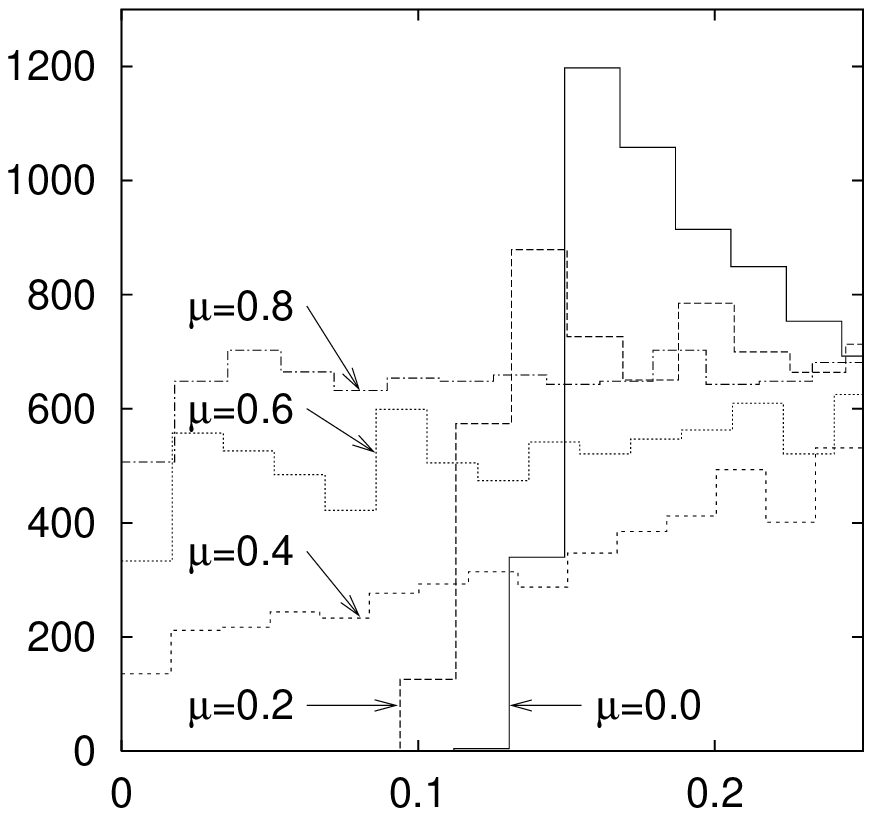,width=4.9cm,height=3.8cm}\hspace*{15mm}
     \epsfig{figure=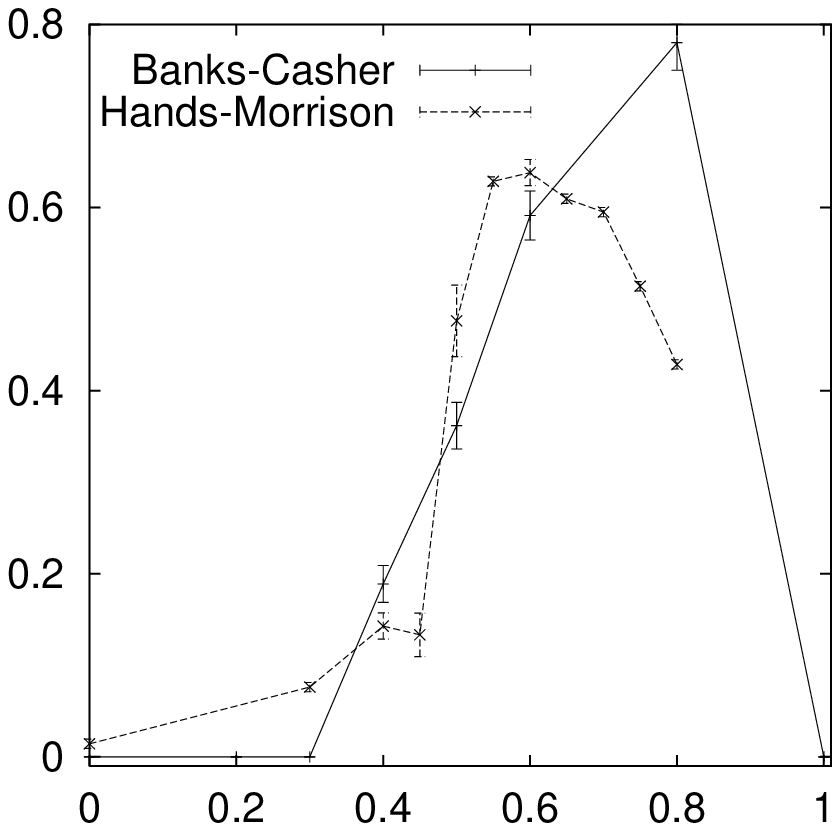,width=4.9cm,height=3.8cm}\\[-42mm]
     \hspace*{-50mm}\phantom{$\rho(y)$}\hspace*{50mm}\phantom{$\langle\psi\psi\rangle$}\\[0.5mm]
     \hspace*{-54mm}$\rho(y)$\hspace*{59mm}$\langle\psi\psi\rangle$\\[28mm]
     \hspace*{50mm}$y$\hspace*{68mm}$\mu$
   \end{center}
\vspace*{-8mm}
   \caption{Upper plots: Density $\rho(y)$ of small eigenvalues of the Dirac operator
        for two-color QCD on a $6^4$ lattice from $\mu=0$ to 0.4 (left).
        The loss of quasi-zero modes is accompanied by a vanishing
        of the chiral condensate. Chiral condensate extracted by different
        methods (right).
            Lower plots: Similar for the ``Gor'kov operator''.}
   \label{fig1}
\end{figure*}

\newabstract 
\begin{center}
{\large\bf Baryons in Partially Quenched and Quenched QCD}\\[0.5cm]
Martin J. Savage\\[0.3cm]
Department of Physics, University of Washington\\
Seattle, Washington, USA
\end{center}

At this point in time it is not possible to perform lattice simulations
of hadronic matrix elements at the physical values of the light quark masses, 
$m_q$.
In order to make some connection between lattice simulations and nature
an extrapolation in the quark masses from those used in the simulations down to
their physical values is required.
Many of us at this meeting have spent a significant fraction of our lives
computing the $m_q$-dependence of hadronic observables in chiral
perturbation theory ($\chi$PT).  However, the results of these calculations can
only be used to extrapolate the results of unquenched simulations, and
unfortunately, these simulations are performed at sufficiently large values of
$m_q$ that the convergence of the chiral expansion is questionable, and most
probably absent.

During the past decade significant progress has been made in understanding how
to make QCD predictions from partially quenched~\cite{Pqqcd} simulations.
The global flavor symmetry group structure 
for partially quenched QCD (PQQCD) with three valence quarks, three ghost
quarks and three sea-quarks, in the chiral limit is
$SU(6|3)_L\otimes SU(6|3)_R\otimes U(1)$ where $SU(m|n)$ is a graded
lie-algebra.
The lowest-lying octet
baryons are included into this framework by embedding them into 
a {\bf 240} dimensional
irreducible representation of $SU(6|3)$~\citethree{LS}{Sa}{CS}, 
and their properties are computed 
systematically in an expansion of $m_{\rm val}/\Lambda_\chi$,  
$m_{\rm sea}/\Lambda_\chi$ and $p/\Lambda_\chi$.
One interesting aspect of this theory is that the extension of electromagnetic,
and weak charges from QCD to PQQCD is not unique.  Different choices correspond
to different weighting's of disconnected quark diagrams.
At the meeting I presented results obtained by Chen and myself~\cite{CS}
for the baryon masses, magnetic
moments and matrix elements of the isovector twist-2 operators.

\newabstract 
\begin{center}
  {\large\bf Spectrum of the QCD Dirac operator in a finite
    volume}\\[0.5cm] 
  Tilo Wettig\\[0.3cm]
  Department of Physics, Yale University, New Haven, CT 06520-8120\\
  and RIKEN-BNL Research Center, Upton, NY 11973-5000\\[0.3cm]
\end{center}

Many hadronic properties depend sensitively on the spectrum of the QCD
Dirac operator which we would like to compute in a finite volume $V$.
In this case, three scales are important: the magnitude of the
smallest Dirac eigenvalues, $\lambda_{\rm min}\sim\pi/V\Sigma$ (with
$\Sigma=|\langle\bar\psi\psi\rangle|$), the Thouless energy, $E_c\sim
f_\pi^2/\Sigma\sqrt{V}$, and the rho mass, $m_\rho$.  Below $m_\rho$,
QCD can be described by an effective chiral theory.  For energies
below $E_c$, the kinetic terms in the effective Lagrangian can be
neglected, but the zero-momentum modes of the Goldstone fields must be
treated nonperturbatively \cite{LS}. This is equivalent to chiral
random matrix theory \cite{VW}.

\begin{figure}[h]
  \begin{center}
    \vspace*{3mm} \epsfig{figure=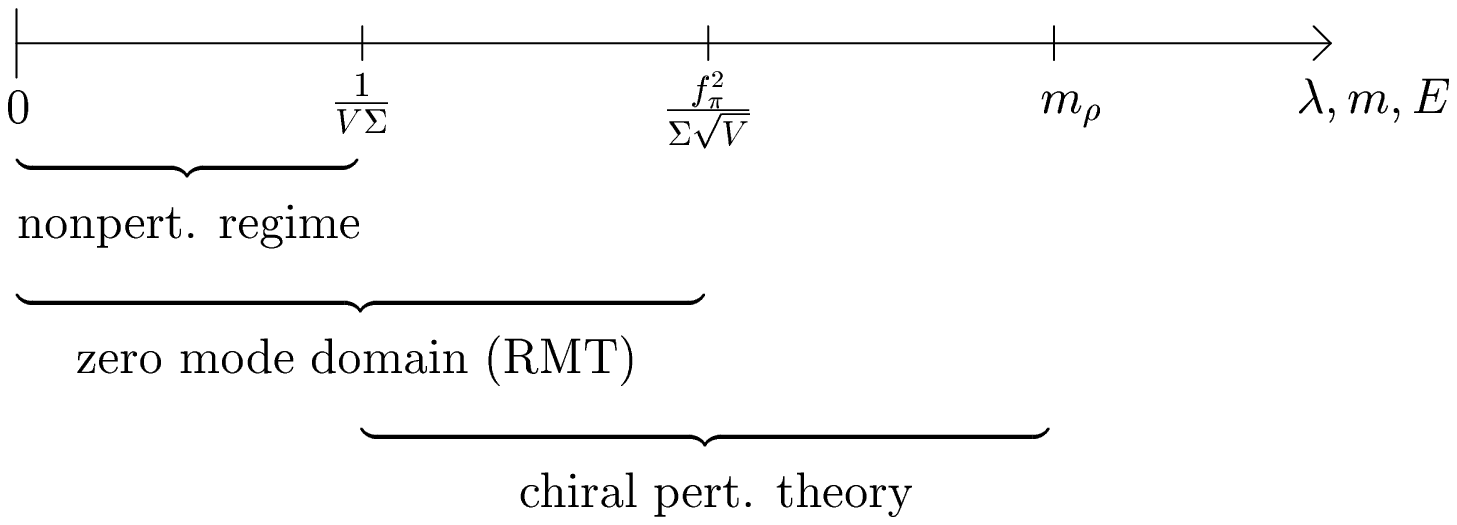,width=60mm} \hspace*{5mm}
    \parbox[b]{75mm}{Fig.~1.  Applicability of chiral RMT and standard
      chiral perturbation theory in a finite volume.  Note the overlap
      region of the two theories, also visible in Fig.~2.}
    \vspace*{-3mm}
  \end{center}
\end{figure} 

We have computed a number of quantities in chiral RMT, chiral
perturbation theory and lattice simulations to confirm this
theoretical picture \cite{GHRSW}, see Fig.~2.

\begin{figure}[h]
  \begin{center}
    \vspace*{3mm} \epsfig{figure=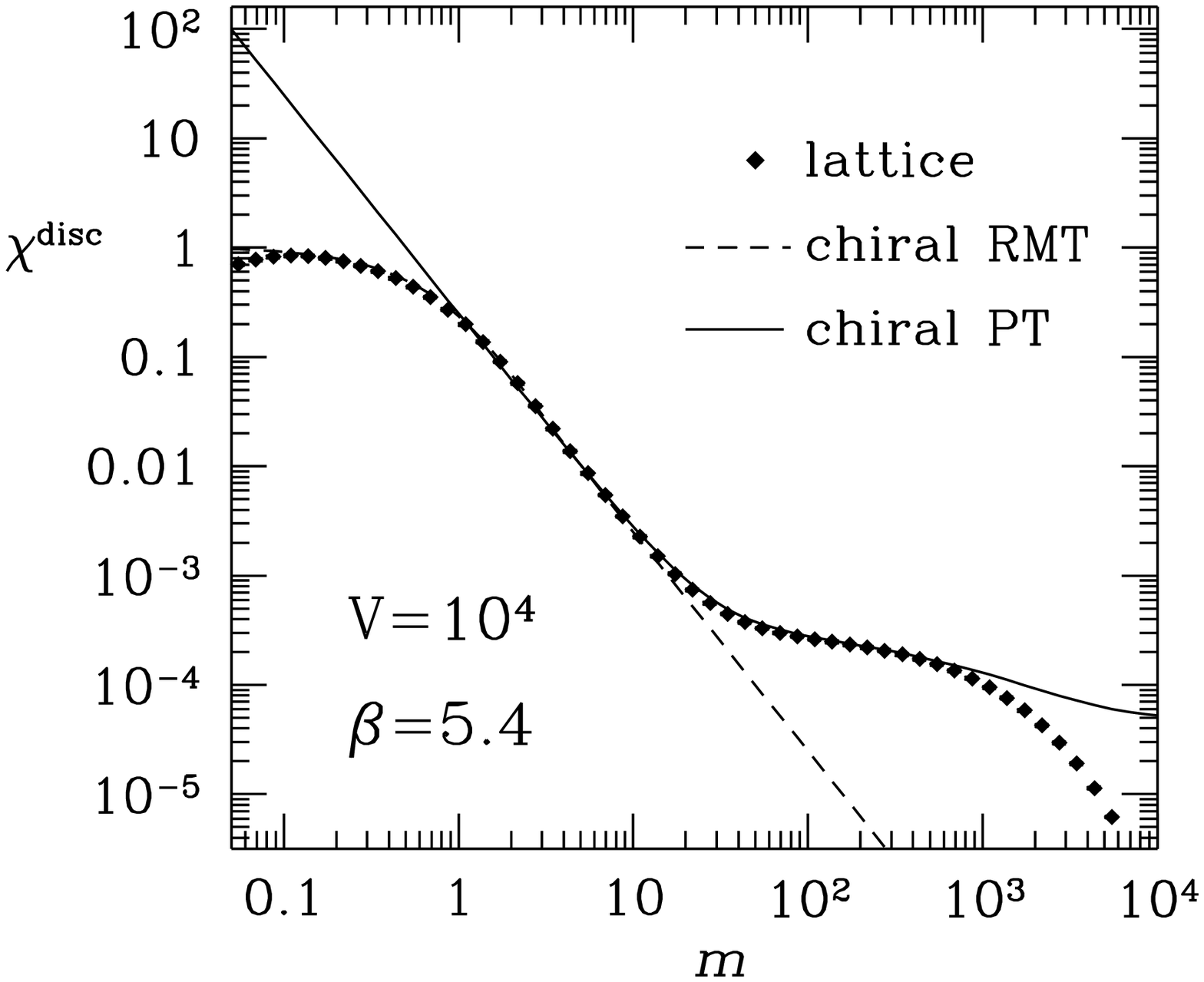,width=60mm} \hspace*{5mm}
    \parbox[b]{75mm}{Fig.~2.  The disconnected scalar susceptibility
      as a function of the valence quark mass from a quenched SU(3)
      simulation using the staggered Dirac operator.  The theoretical
      curves are from chiral RMT and from a finite-volume chiral
      perturbation theory applicable for staggered fermions at
      moderate to strong coupling, see Ref.~\cite{GHRSW}.  Our
      calculations nicely confirm the theoretical expectation of
      Fig.~2.}
    \vspace*{-3mm}
  \end{center}
\end{figure}

I would like to thank M. G\"ockeler, H. Hehl, P.E.L. Rakow, and A.
Sch\"afer for a fruitful collaboration on these topics.

\newabstract 

\begin{center}
{\large\bf Topological lumps near the QCD phase transition}\\[0.5cm]
Christof Gattringer\\[0.3cm]
Institut f\"ur Theoretische Physik, Universit\"at Regensburg, 
D-93040 Regensburg.\\[0.3cm]
\end{center}

During the last 20 years a quite comprehensive picture of instantons 
in QCD has been developed \cite{SchSh98}. Of particular 
importance is the understanding of chiral symmetry 
breaking: A fluid of weakly interacting instantons produces a non-vanishing 
density of eigenvalues near the origin. This spectral density is 
related to the chiral condensate through the Banks-Casher relation 
\cite{BaCa80}. 
When increasing the temperature above its critical value $T_c$
chiral symmetry is 
restored. At this transition the topological lumps 
are expected to form tightly bound molecules 
of instantons and anti-instantons. These molecules do no longer produce
eigenvalues near the origin, the spectrum develops a gap and chiral symmetry
is restored. 

In a series of articles \cite{Gaetal1} we analyzed evidence for chiral 
symmetry breaking through instantons from ab-initio calculations on the
lattice. We studied eigenvalues and eigenvectors of 
an approximate Ginsparg-Wilson fermion \cite{GiWi82} developed in 
\cite{Gaetal2}. The eigenvectors are known to trace the topological 
lumps of the underlying gauge fields and thus can be used to analyze the 
instanton structure. In particular we focussed on the localization and 
local chirality \cite{locchir} of the eigenvectors.

We show that the most localized states can be found near the origin 
($T < T_c$) respectively near the edge of the spectral gap ($T > T_c$). 
Below $T_c$ the eigenstates
are chiral, i.e.~the underlying instantons are 
relatively unperturbed. Above $T_c$ the states have lost their
local chirality and one can no longer speak of intact instantons. Our findings
support the picture of chiral symmetry breaking through instantons and
the mechanism for chiral symmetry restoration above $T_c$.

\newabstract 

\begin{center}
{\large\bf Direct Determination of Flavor Singlet Masses from the Lattice}\\[0.5cm]
{\bf K. Schilling}$^a$, H. Neff$^a$, N. Eicker$^a$, J. Negele$^b$
and Th. Lippert$^a$
\\[0.3cm]
Bergische Universitat Wuppertal,\\
    Fachbereich 8, Physik\\
    Gausstrasse 20, 42097 Wuppertal,    Germany\\[0.3cm]
$^b$Massachusetts Institute of Technology\\
    Lab for Nuclear Science \& Center for Theoretical Physics, Room 6-113,\\
 77 Massachusetts Ave.
    Cambridge, Massachusetts 02139-4307\\[0.3cm]
\end{center}

The $\eta'$-mass is believed to be related to the $U_A(1)$
anomaly and hence bears important witness of the topological 
structure of the QCD vacuum, as suggested by the Witten-Veneziano
formula that  holds in the large $N_c$ limit of the theory. So far, flavor
singlet mesons have been rather elusive to the lattice approach 
because of large fluctuations encountered in the so-called OZI-rule suppressed
diagrams.

We present first evidence\citetwo{1}{2}
from two-flavor QCD for a {\it direct signal} of
such flavor singlet meson masses, by applying
\begin{itemize}
\item
spectral methods to  the OZI-rule
forbidden piece and
\item
ground state projection techniques to  the
OZI-rule allowed contribution 
\end{itemize}
of the singlet meson propagator.

\newabstract 

\begin{center}
{\large\bf Calculating $K\rightarrow\pi\pi$ decay amplitudes beyond the
leading-order chiral expansion}\\[0.5cm]
C.-J. David Lin$^1$ \\[0.3cm]
$^1$Department of Physics and Astronomy,\\
The University of Southampton,\\
Southampton SO17 2JF,\\ENGLAND\\[0.3cm]
\end{center}

The calculation of non-leptonic decays of hadrons from lattice QCD presents
a very challenging problem, mainly because of the formulation of the
theory in Euclidean space \cite{Maiani:1990ca}.
The finite-volume techniques developed in Refs. 
\citetwo{Lellouc:2000pv}{Lin:2001ek} offer a possibility of tackling 
$K\rightarrow\pi\pi$ decays.  These techniques do not rely on any 
expansion, and yield results of exponential precision in the volume,
hence they will eventually give very accurate estimates for 
$\langle K | {\mathcal{O}} | \pi\pi\rangle$.  Nevertheless, implementing
these finite-volume techniques is computationally demanding.

\medskip

Traditionally, lattice calculations for $K\rightarrow\pi\pi$ amplitudes
have been done, following the
proposal in \cite{Bernard:1985wf}, by computing matrix elements
of the kind $\langle K|{\mathcal{O}}|\pi\rangle$ then relating them 
to $\langle K|{\mathcal{O}}|\pi\pi\rangle$ via the lowest-order
chiral expansion.  In this approach, the chiral corrections are
completely missing, and it is very difficult to study these corrections.

\medskip

In this talk, a novel approach to the calculation of $K\rightarrow\pi\pi$
amplitudes at the next-to-leading order chiral expansion is presented.
The basic idea is to compute matrix elements
of the kind $\langle K|{\mathcal{O}}|\pi\pi\rangle$ directly on the 
lattice, allowing energy-momentum injections via the weak operator.
More details can be found in Refs. \citetwo{Boucaud:2001mg}{Boucaud:2001tx}.

\end{document}